\documentclass[fleqn,usenatbib]{mnras}

\usepackage[T1]{fontenc}
\usepackage{ae,aecompl}

\usepackage{graphicx}	
\usepackage{amsmath}	
\usepackage{amssymb}	

\usepackage{epstopdf}

\newcommand{\Ha}{\ensuremath{\mathrm{H} \alpha}}
\newcommand{\Hb}{\ensuremath{\mathrm{H} \beta}}
\newcommand{\Hi}{H~{\sc i}}
\newcommand{\Hii}{H~{\sc ii}}

\newcommand{\nii}{\ensuremath{\textrm{[N~{\sc ii}]}}}

\newcommand{\cm}{\ensuremath{\textrm{ cm}}}

\newcommand{\cucm}{\ensuremath{\textrm{ cm}^{-3}}}

\newcommand{\radmsq}{\ensuremath{\textrm{ rad m}^{-2}}}
\newcommand{\kms}{\ensuremath{\textrm{ km s}^{-1}}}
\newcommand{\kpc}{\ensuremath{\, \mathrm{kpc}}}
\newcommand{\pc}{\ensuremath{\textrm{ pc}}}

\newcommand{\K}{\ensuremath{\textrm{ K}}}

\newcommand{\R}{\ensuremath{\textrm{ R}}}
\newcommand{\uG}{\ensuremath{\, \mu \mathrm{G}}}

\newcommand{\ncre}{\ensuremath{n_\mathrm{cre}}}

\newcommand{\MHz}{\ensuremath{\textrm{ MHz}}}
\newcommand{\GHz}{\ensuremath{\textrm{ GHz}}}

\newcommand{\vlsr}{\ensuremath{V_\mathrm{LSR}}}

\newcommand{\pitch}{\ensuremath{\Theta}}

\renewcommand{\vec}{\mathbf}
\newcommand{\arcdeg}{\ensuremath{^\circ}}

\defcitealias{GaoReich:2015}{GRR15}

\title[The Fan Region extends beyond the Perseus Arm]{The Fan Region at 1.5 GHz. I: Polarized synchrotron emission extending beyond the Perseus Arm}

\author[A. S. Hill et al.]{A. S. Hill,$^{1,2}$\thanks{Email:
\href{mailto:ashill@haverford.edu}{ashill@haverford.edu}} T. L. Landecker,$^3$ E. Carretti,$^{1,4}$ K. Douglas,$^{5}$
X. H. Sun,$^{6}$ 
\newauthor B. M. Gaensler,$^{6,7}$ S. A. Mao,$^{8}$ N. M.
McClure-Griffiths,$^{1,9}$ W. Reich,$^8$
\newauthor M. Wolleben,$^{10}$ J. M. Dickey,$^{11}$ A. D. Gray,$^{3}$ M. Haverkorn,$^{12}$ J. P. Leahy,$^{13}$ and
\newauthor D. H. F. M. Schnitzeler$^{8}$
\\
$^1$CSIRO Astronomy \& Space Science, PO Box 76, Epping, NSW 1710, Australia \\
$^2$Department of Astronomy, Haverford College, Haverford, PA 19041, USA \\ 
$^3$National Research Council Canada, Herzberg Program in Astronomy and
Astrophysics, \\ Dominion Radio Astrophysical Observatory, PO Box 248,
Penticton, British Columbia V2A~6J9, Canada \\
$^4$Present address: INAF/Osservatorio Astronomico di Cagliari, Via della Scienza 5, I-09047
Selargius, Italy\\
$^5$Physics and Astronomy Department, Okanagan College, 1000 KLO Road, Kelowna,
British Columbia V1Y~4X8, Canada\\
$^6$Sydney Institute for Astronomy, School of Physics, University of Sydney,
NSW 2006, Australia \\
$^7$Dunlap Institute for Astronomy and Astrophysics, University of Toronto,  50
St. George Street, Toronto, Ontario M5S~3H4, Canada \\
$^8$Max-Planck-Institut f\"ur Radioastronomie, Auf dem H\"ugel 69, D-53121,
Bonn, Germany \\
$^9$Research School of Astronomy and Astrophysics, Australian National University, Canberra, ACT 2611, Australia \\
$^{10}$Skaha Remote Sensing, 3165 Juniper Drive, Naramata, BC V0H 1N0,
Canada \\
$^{11}$University of Tasmania, School of Mathematics and Physics, Hobart, TAS 7001, Australia \\
$^{12}$Department of Astrophysics/IMAPP, Radboud University Nijmegen, PO Box
9010, NL-6500 GL, Nijmegen, the Netherlands \\
$^{13}$Jodrell Bank Centre for Astrophysics, Alan Turing Building, School of
Physics and Astronomy, University of Manchester,\\ Oxford Road, Manchester
M13~9PL, UK 
}

\date{Accepted XXX. Received YYY; in original form ZZZ}

\volume{in press}
\pubyear{2017}

\begin{document}
\label{firstpage}
\pagerange{\pageref{firstpage}--\pageref{lastpage}}
\maketitle

\begin{abstract}
The Fan Region is one of the dominant features in the polarized radio sky, long
thought to be a local (distance $\lesssim 500 \pc$) synchrotron
feature. We present $1.3-1.8 \GHz$ polarized radio continuum observations of the region from
the Global Magneto-Ionic Medium Survey (GMIMS) 
and compare them to maps of \Ha\ and polarized radio continuum intensity from $0.408-353 \GHz$.
The high-frequency ($> 1 \GHz$) and low-frequency
($\lesssim 600 \MHz$) emission have different morphologies, suggesting a different
physical origin. Portions of the $1.5 \GHz$ Fan Region emission are
depolarized  by $\approx 30\%$ by ionized gas structures in the Perseus Arm,
indicating that  this fraction of the emission originates $\gtrapprox 2 \kpc$ away.
We argue for the same conclusion based on the high polarization fraction at
$1.5 \GHz$ ($\approx 40\%$).
The Fan Region is offset with respect to the Galactic plane, covering
$-5\arcdeg \lesssim b \lesssim +10\arcdeg$; we attribute this offset to the
warp in the outer Galaxy.
We discuss origins of the polarized emission, including the spiral
Galactic magnetic field. This idea is a plausible contributing factor
although no model to date readily reproduces all of the observations.
We conclude that models of the Galactic
magnetic field should account for the $\gtrsim 1 \GHz$ emission from the Fan Region
as a Galactic-scale, not purely local, feature.

\end{abstract}




\section{Introduction}

Linearly polarized radio continuum emission arises from the interaction of
ultrarelativistic electrons with a magnetic field. The first detections of
polarized emission from the Galaxy
\citep{WesterhoutSeeger:1962,WielebinskiShakeshaft:1962} already recognized the
two large features that dominate the northern polarized sky, the North Polar
Spur and the Fan Region. The Fan Region extends over a $\sim 60^\circ \times 30^\circ$
region, centred at Galactic longitude $\ell \approx 130\arcdeg$
slightly above the Galactic plane at Galactic latitude $b \approx +5^\circ$. The
Fan Region is remarkable for the intensity of its polarized radiation and the
regularity of its polarization angle. In total intensity images,
it does not stand out from its surroundings.
It is identified by (and named for) electric field vectors which
appear to fan out from the Galactic plane near $\ell=130^\circ$ at low frequencies
$\nu \lesssim 600 \MHz$ \citep{BinghamShakeshaft:1967,BrouwSpoelstra:1976}.
Polarized emission in this region of the sky is evident from $\approx 100 \MHz$
\citep{IacobelliHaverkorn:2013a} to $353 \GHz$ \citep{Planck-CollaborationAde:2015}.
All of this emission is generally referred to as the Fan Region.

The origin of the Fan Region is unknown, but most authors have considered it a
local ($d \lesssim 500 \pc$) feature \citep{WilkinsonSmith:1974,Spoelstra:1984uu}.
\citet{Verschuur:1968wh} identified a depolarized
ring feature at $408 \MHz$ at $(137\arcdeg,+7\arcdeg)$\footnote{We denote positions in Galactic
coordinates as $(\ell, b)$.} which they associated with a
star $140-200 \pc$ from the Sun to establish a lower limit to the distance;
\citet{IacobelliHaverkorn:2013a} placed this ring $\approx 200 \pc$ away based on
$150-350 \MHz$ observations.
\citet{WilkinsonSmith:1974} found no depolarization at $\nu \le 610\MHz$ due to the \Hii\
region Sh2-202, establishing
an upper limit. The modern distance to Sh2-202 is $0.97 \pm 0.08 \kpc$
\citep{FosterBrunt:2015}.
These arguments for a local origin of the Fan Region are based primarily on
low-frequency observations. 
In contrast, \citet{BinghamShakeshaft:1967} argued that the high polarization
fraction at $1407 \MHz$ can only be produced by Galactic structure.
\citet{Wolleben:2005} found depolarization by numerous \Hii\ regions and argued
that the $1.4 \GHz$ emission
occurs over a range of distances from $\approx 500 \pc$ to a few kpc, a range
which includes both local gas and the Perseus spiral arm. Because the intrinsic
polarization angle of synchrotron radiation is related to the orientation on the
sky of the magnetic field in the emitting region, if the Fan Region emission
originates over this long path length, it must indicate a uniform Galactic
magnetic field on kpc scales in this direction \citep{WollebenLandecker:2006}. 

In this paper, we present $1.5 \GHz$ polarized continuum observations of the Fan
Region.
Our focus here is on morphological comparisons between continuum observations
from $0.4$ to $353 \GHz$ and with spectroscopically-resolved \Ha\ observations.
We present our data in Section~\ref{sec:data}, briefly reviewing relevant
depolarization mechanisms in Section~\ref{sec:emission}. We discuss the kinematic features
seen in \Ha\ and \Hi\ observations in the direction of the Fan Region as they
relate to Galactic structure in Section~\ref{sec:features}. We describe the Fan
Region at all wavelengths and compare the morphology to observations of interstellar
medium (ISM) structures with known distances in Section~\ref{sec:results}.
In Section~\ref{sec:polfrac}, we discuss the implications of the high observed
fractional polarization in the Fan Region. In Sections~\ref{sec:emis} and
\ref{sec:structure}, we construct a simple model of the synchrotron emission due
to Galactic spiral structure, incorporating the effects of geometrical and depth
depolarization, and compare the results to the observed synchrotron intensity as
a function of longitude.
We summarize the paper and draw conclusions in Section~\ref{sec:conclusions}.
In Paper~II (A.~S.\ Hill et al in prep), we will model depolarization due to
Faraday effects in the Fan Region.

\section{Observations} 
\label{sec:data}

\subsection{GMIMS survey} \label{sec:gmims_data}

We use radio polarization data from the Global Magneto-Ionic Medium Survey
(GMIMS) high-band north (GMIMS-HBN). In GMIMS \citep{WollebenLandecker:2009},
we are using telescopes around
the world to map polarized emission from the entire sky, north and south,
spanning 300 to 1800~MHz. The survey is designed to measure the polarized
intensity, $L(\phi)$, as a function of Faraday depth
\begin{equation} \label{eq:phi}
\phi(s) = K \int^{\mathrm{observer}}_{s} n_e(s') \, \vec{B}(s') \cdot d\vec{s'}.
\end{equation}
Here $K \equiv e^3 / (2 \pi m_e^2 c^4) = 0.81 \, (\mathrm{cm}^{-3} \uG \pc)^{-1}
\radmsq$, $n_e$ is the electron density in the intervening ISM, and $\vec{B}$
is the magnetic field. We acquire data in thousands of frequency channels
to allow us to use 
rotation measure (RM) synthesis \citep{BrentjensBruyn:2005}.

The GMIMS-HBN data were acquired with the 26~m John A.\ Galt Telescope at the
Dominion Radio Astrophysical Observatory (DRAO) with continuous frequency
coverage from 1280 to 1750~MHz. Data were acquired in 2048 individual channels of
width 236.8~kHz. \citet{WollebenFletcher:2010} describe the receiver and the data acquisition
process for GMIMS-HBN, and \citet{WollebenLandecker:2010} and \citet{SunLandecker:2015}
give examples of use of GMIMS-HBN data. The full survey will be presented and
publicly released elsewhere.

The observations were made by moving the telescope slowly up and down the meridian
as the sky moved by; each such telescope track is referred to as a ``scan''. Earth
rotation during a scan caused each scan to follow a diagonal track across the equatorial
coordinate grid. Successive up and down scans were made until the sky was fully sampled
between declinations $-30\arcdeg$ and $+87\arcdeg$. After calibration, the
many scan crossings were reconciled using the ``basketweaving'' technique
\citep{WollebenLandecker:2010}, which we used to iteratively deduce the best-fit
zero level for each scan. This process strips the sky minimum, which makes the zero
point for Stokes $I$ measurements inconsistent \citep{WollebenLandecker:2010}, so
we do not use total intensity data from GMIMS-HBN. Observations were
made between sunset and sunrise to avoid contamination through sidelobes by radio emission
from the Sun. The angular resolution varies from $40'$ to $30'$ across this frequency
range; we have smoothed the data to a common resolution of $40'$ and reprojected
to a plate carr\'ee projection \citep{CalabrettaGreisen:2002}.

Daily calibration observations were made of the bright small-diameter sources
Cas~A, Cyg~A, Tau~A, and Vir~A. Using flux densities and spectral indices of
these sources from \citet{BaarsGenzel:1977}, we converted the scan data to
units of Janskys. The conversion factor from Janksys to Kelvins of main beam antenna
temperature (equivalent to the gain of the telescope) was established from
careful measurement of the antenna temperature produced by Cyg~A on an absolute
temperature scale established with resistive terminations at liquid nitrogen
temperature and at $\approx 100\arcdeg \, \mathrm{C}$. Finally, data were converted to
main-beam brightness temperatures by dividing by the beam efficiency of the
telescope. We consider the temperature scale to be correct within $\sim{3}\%$.
\citet{DuLandecker:2016} present details of the determination of telescope gain.

The Stokes $Q$ and $U$ spectra are smooth and there is no evidence of bandwidth
depolarization in individual channels.
\citet{SunLandecker:2015} made tests of data quality from the GMIMS data in the vicinity of the North Polar Spur, another region of bright polarized emission. They concluded that the data set is of high quality in regions of bright polarized emission. 

We apply RM synthesis to our polarization data cubes.
In RM synthesis, we construct the Faraday dispersion function, which takes the form
of a Fourier transform of the observed complex polarization vector, $\mathcal{P} = Q + iU$.
The Faraday dispersion function is integrated over the interference-free portions
of the observed band. At a given Faraday depth, RM synthesis accounts for rotation
in the polarization angle over the band.
Due to Faraday rotation, there is no single polarization angle which describes the
data at all frequencies in the band. Without RM synthesis, integrating over many
channels would lead to significant bandwidth depolarization
\citep[e.g.][]{BrentjensBruyn:2005,Heald:2009kc,SchnitzelerKatgert:2009}.

RM synthesis also allows us to separate the emission as a function of $\phi$.
For each pixel on the sky, we construct the polarized intensity $L_{1.5}(\phi)$ sampled every $5 \radmsq$.
Our observing frequencies and spectral resolution leave us sensitive to emission with
$|\phi| < 2 \times 10^5 \radmsq$; the resolution is $\delta \phi = 149 \radmsq$
\citep[calculated following][]{SchnitzelerKatgert:2009}.
The data were recorded and Faraday depth spectra
calculated in equatorial coordinates; we subsequently reprojected to Galactic coordinates.
From each Faraday depth spectrum, we calculated the peak polarized intensity at each pixel using
a three-point quadratic fit with the {\tt miriad} task {\tt moment}. The polarized
intensity images we present in this paper are images of this peak polarized
intensity. The Faraday depth of the peak is typically at $|\phi| < 10 \radmsq$,
so these images are similar to an image of the $\phi = 0 \radmsq$ channel. With
the large $\delta \phi$, we assume that there is only a single component resolved
by the GMIMS-HBN observations.

When all of the polarized signal is at a single Faraday depth, an image produced
with RM synthesis has the noise expected from the entire band rather than the noise
from individual frequency channels.
The noise in each $\phi$ channel is $\sigma_L \approx 0.02 \K$, making the signal-to-noise
ratio in the Fan Region $\gtrsim 20$ on a single-pixel basis. This allows us to
measure the centroid of Faraday depth components with an uncertainty of
$\approx 3 \radmsq$ in the Fan Region.
The structures in Faraday depth in the Fan Region seen
in low-frequency observations are typically
$\sim 1-10 \radmsq$ in extent \citep{IacobelliHaverkorn:2013}, so we do not resolve
multiple Faraday depth components with the GMIMS-HBN data; a future low-frequency
component of the GMIMS survey will enable the separation of these narrow Faraday
depth components. With a maximum frequency of $1750 \MHz$, we are not sensitive to
individual Faraday depth features which are wider than $\approx 107 \radmsq$
\citep{BrentjensBruyn:2005}.

\citet{WollebenLandecker:2006} presented an absolutely-calibrated $1.4 \GHz$
polarization survey also using John A.\ Galt Telescope observations. This older survey
employed a single channel of bandwidth $12 \MHz$ and a drift scanning strategy. Each
drift scan was Nyquist-sampled in right ascension but the survey achieved
$41.7\%$ of Nyquist sampling in declination. The GMIMS-HBN survey improves upon
the \citet{WollebenLandecker:2006} survey with a much wider bandwidth and full
Nyquist sampling. Moreover, with the basketweaving observing strategy, each point
is observed twice, reducing uncertainty relative to the drift scan strategy.

\subsection{Complementary data sets}

\begin{table*}
 \caption{Data sets used. Bandwidths and channel widths are listed in frequency
 units for continuum surveys and velocity units for spectral line surveys. The
 sampling column lists the approximate typical beam spacing on the sky in the
 midplane for surveys which are not Nyquist sampled. 
 {\em References}: Dwingeloo: \citet{BrouwSpoelstra:1976,CarrettiBernardi:2005}.
 Stockert/Villa Elisa: \citet{Reich:1982va,ReichReich:1986,ReichTestori:2001}.
 CGPS: \citet{TaylorGibson:2003,LandeckerReich:2010}.
 LAB: \citet{KalberlaBurton:2005}.
 GMIMS-HBN: This work.
 Urumqi: \citet{GaoReich:2010}.
 {\em WMAP}: \citet{BennettLarson:2013}.
 {\em Planck}: \citet{TauberMandolesi:2010,Planck-CollaborationAdam:2016}.
 WHAM-SS: \citet{HaffnerReynolds:2003,HaffnerReynolds:2010}.
 }
 \label{tbl:data}
 \begin{tabular}{llrrrr@{.}lcl}
  \hline
  Survey & Stokes/ & \multicolumn{1}{c}{Frequency/} & Bandwidth & \multicolumn{1}{c}{Channel} & \multicolumn{2}{c}{Beam} & Sampling & Coverage \\
         & Line & \multicolumn{1}{c}{Wavelength} &           & \multicolumn{1}{c}{width} & \multicolumn{2}{c}{FWHM} &          &          \\
  \hline
  Dwingeloo & $Q,U$       & $408 \MHz$ & $2 \MHz$   & $2 \MHz$   & 2&3\arcdeg  & $2.3\arcdeg$ (irregular) & Northern sky \\
  Dwingeloo & $Q,U$       & $465 \MHz$ & $2 \MHz$   & $2 \MHz$   & 2&0\arcdeg  & $2.3\arcdeg$ (irregular) & Northern sky \\
  Dwingeloo & $Q,U$       & $610 \MHz$ & $4 \MHz$   & $4 \MHz$   & 1&5\arcdeg  & $2.3\arcdeg$ (irregular) & Northern sky \\
  Dwingeloo & $Q,U$       & $820 \MHz$ & $4 \MHz$   & $4 \MHz$   & 1&0\arcdeg  & $2.3\arcdeg$ (irregular) & Northern sky \\
  Dwingeloo & $Q,U$       & $1411 \MHz$ & $7 \MHz$  & $7 \MHz$   & 0&6\arcdeg  & $2.3\arcdeg$ (irregular) & Northern sky \\
  Stockert/Villa Elisa  & $I$ & $1420 \MHz$ & $18 \MHz$ & $18 \MHz$ & 0&6\arcdeg  & Nyquist & All sky \\
  CGPS      & $I,Q,U$     & $1420 \MHz$ & $35 \MHz$ & $35 \MHz$  & 0&02\arcdeg          & Nyquist & $-3\arcdeg < b < +5\arcdeg$ \\
  LAB       & \Hi  & $1421 \MHz$ & $\pm 400 \kms$   & $1 \kms$   & 0&6\arcdeg & $0.5\arcdeg$ & All sky \\
  GMIMS-HBN & $Q,U$       & $1500 \MHz$ & $470 \MHz$ & $0.2 \MHz$ & 0&6\arcdeg  & Nyquist & $-30\arcdeg < \delta < +87\arcdeg$ \\
  Urumqi    & $I, Q, U$   & $4800 \MHz$ & $600 \MHz$ & $600 \MHz$ & 0&16\arcdeg & Nyquist & W4 superbubble \\
  {\em WMAP}      & $I,Q,U$     & $22.8 \GHz$ & $5.5 \GHz$ & $5.5 \GHz$ & 0&82\arcdeg  & Nyquist & All sky \\
  {\em Planck}    & $I,Q,U$     & $353 \GHz$ & $116 \GHz$ & $116 \GHz$ & 0&08\arcdeg  & Nyquist & All sky \\
  WHAM-SS  & \Ha         & $656.3$~nm & $\pm 80 \kms$ & $12 \kms$ & 1&0\arcdeg  & $1.0\arcdeg$ & All sky \\
  \hline
 \end{tabular}
\end{table*}

In addition to the GMIMS data presented here, we use several published data sets
which provide complementary information. To trace synchrotron emission adequately,
we have chosen data sets at a range of frequencies because depolarization cannot be
understood from one frequency alone. We have also chosen data sets which trace the dust
and ionized and neutral gas (with spectral resolution allowing separation of
emission due to Galactic rotation) in the diffuse ISM.
We have regridded all of the data sets
to a plate carr\'ee projection image with $0.5\arcdeg$ pixels. We list frequencies,
bandwidths, beam sizes, sampling, and the coverage as is relevant to the Fan
Region of each of these surveys in Table~\ref{tbl:data}. We
refer the reader to the references in Table~\ref{tbl:data} for details
but mention the most important points for our work here.

We use spectroscopic maps of \Ha\ emission from data release~1 of the all-sky
Wisconsin H-Alpha Mapper Sky Survey (WHAM-SS)\footnote{\url{http://www.astro.wisc.edu/wham/}}
and of \Hi\ emission from the Leiden-Argentine-Bonn
(LAB) survey. We use the low-frequency continuum surveys of \citet{BrouwSpoelstra:1976},
compiled and resampled on a regularly gridded map by \citet{CarrettiBernardi:2005}.
These observations used rotating feed antennas and thus record the polarized intensity
and polarization angle directly \citep{BerkhuijsenBrouw:1964}. Especially at the
higher frequencies, these data are severely undersampled (see Table~\ref{tbl:data}).
We use $23 \GHz$ data from the nine-year data release of the {\em WMAP} experiment.
The high frequency makes these data virtually free from Faraday rotation effects,
either angle rotation or depolarization.
For the Urumqi $4.8 \GHz$ observations, \citet{GaoReich:2010} set the zero
level by extrapolating the WMAP $23 \GHz$ data using a spectral index measured
from $23 \GHz$ (WMAP) to $1.4 \GHz$ \citep{WollebenLandecker:2006}. This
assumes that there is little Faraday depolarization across the band
\citep{SunHan:2007}, an assumption supported by consistent spectral indices.
However, the \citet{WollebenLandecker:2006} data do not show the depolarization
effects that are evident in the GMIMS-HBN data presented here (see
\S~\ref{sec:morphology} below). If there is any Faraday depolarization at
$4.8 \GHz$ on scales larger than a few degrees (where the scaled $23 \GHz$ data
sets the zero level), the Urumqi $4.8 \GHz$ data would not be sensitive to it.

In the $353 \GHz$ image from the {\em Planck} mission, the thermal emission from dust
dominates over other contributions, and the polarization of the signal traces the
Galactic magnetic field \citep{Planck-CollaborationAde:2015}.
The Canadian Galactic Plane Survey (CGPS) provides high angular resolution
polarized radio continuum data at $1420 \MHz$ for a small portion of the Fan
region using the DRAO Synthesis Telescope. The \citet{WollebenLandecker:2006}
data provide information on the largest angular scales for the CGPS data;
Effelsberg 100~m Telescope data provide information on intermediate scales.

\subsection{Synchroton emission and depolarization} \label{sec:emission}

The polarized radio continuum emission we present in this paper is primarily due to
synchrotron emission and, at $353 \GHz$, dust emission. As polarized radiation
propagates through the ionized ISM, depolarization
effects reduce the polarized intensity by producing polarization vectors at different
angles either along the line of sight or within the beam. The superposition
of these vectors reduces the observed polarized intensity
\citep{Burn:1966ug,Tribble:1991us,SokoloffBykov:1998,GaenslerDickey:2001}. {\em Beam
depolarization} arises when different paths within the beam have different
Faraday depths (see eq.~\ref{eq:phi}) and thus
polarization angles. Foreground turbulent regions can lead to significant beam
depolarization. {\em Depth depolarization} occurs when distant emission is
Faraday-rotated and cancels more local emission. Depth depolarization generally
refers to an effect that occurs over a long path. Faraday screens can rotate
background Faraday rotation and cause depolarization in a similar manner but
over a very short path \citep{SunHan:2007}; we refer to this as {\em Faraday
screen depolarization}.
These three effects are results of Faraday rotation and are thus frequency-dependent.
{\em Geometrical depolarization} is the product of different orientations of the magnetic
field at different distances along the line of sight resulting in the
superposition of emission with different intrinsic polarization angles
\citep{Miville-DeschenesYsard:2008,DelabrouilleBetoule:2013}. Geometrical
depolarization is the most important depolarization mechanism at high
frequencies. At $23 \GHz$, the
change in polarization angle $\Delta \psi \approx 0.01 \textrm{ rad}$ for
$\phi=50 \radmsq$), so Faraday rotation is negligible.

\section{Kinematic features and distances} \label{sec:features}

Because all components of the ISM (atomic, molecular, and ionized gas as well as
the magnetic field) are inter-related, we examine the role of Galactic structure
in producing both the ionized gas and the polarized emission from the Fan Region.
In this section, we review our knowledge of Galactic structure and the distances to observed
features in the second and third quadrants of the Galaxy.

\begin{figure}
\includegraphics[width=0.5\textwidth]{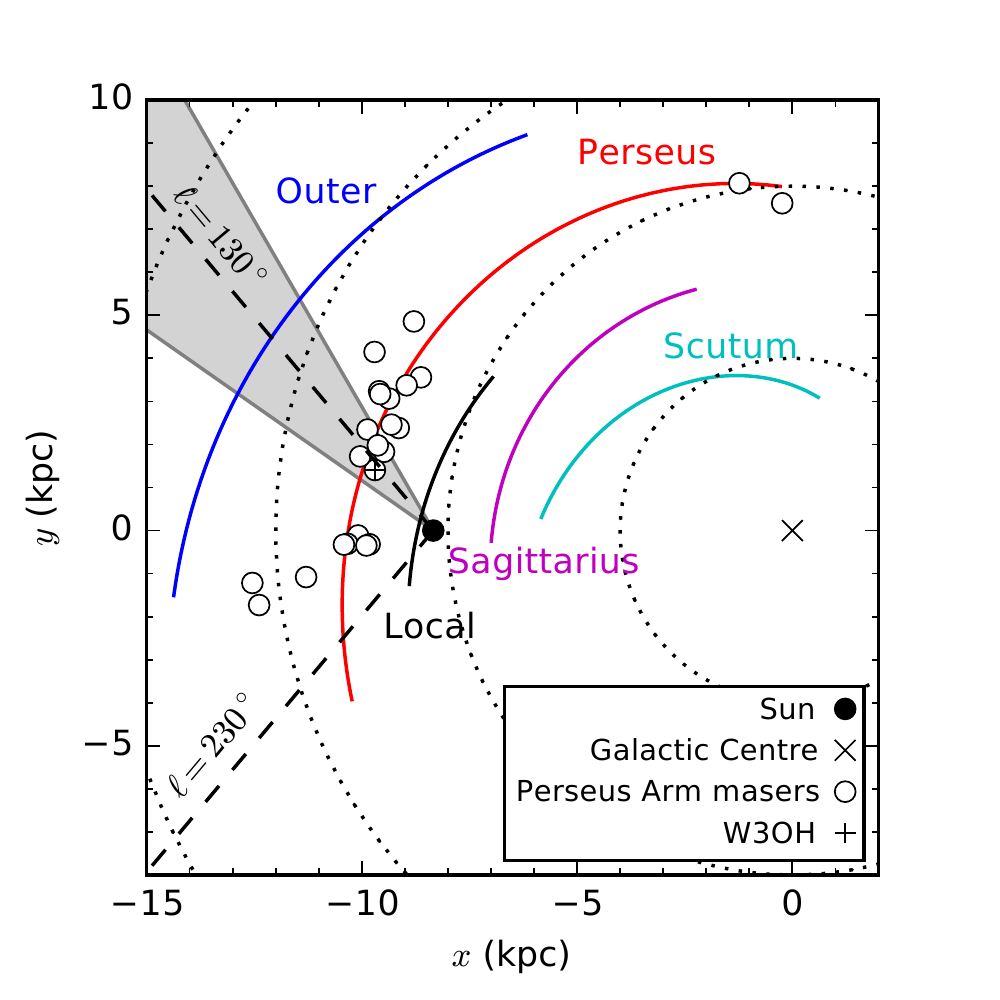}
\caption{View of the Galaxy from above with spiral arms from \citet{ReidMenten:2014}.
The longitude range $120\arcdeg < \ell < 145\arcdeg$, the region which contains the brightest emission from the Fan Region, is shaded gray. 
}
\label{fig:arms}
\end{figure}

Sightlines towards the Fan Region ($\ell \approx 130\arcdeg$) pass through the Perseus
and outer spiral arms. We illustrate this with an overhead view of the Galaxy in
Figure~\ref{fig:arms}. The Perseus Arm, which \citet{Benjamin:2008uh} considers
one of two ``major'' spiral arms, is observed in ionized and neutral gas and
star formation. It is about $2 \kpc$ away at a Galactocentric radius of $\approx 10
\kpc$, while the Outer Arm, likely a more minor arm with a
concentration of gas but not of old stars, is $\approx 6 \kpc$ away at a
Galactocentric radius of $\approx 13 \kpc$
\citep[e.g.][]{XuReid:2006,ChurchwellBabler:2009,ReidMenten:2014}.

In the second quadrant, sightlines are closer to normal to these spiral arms than
are similar sightlines in the third quadrant. As an example, we consider the two
sightlines $50\arcdeg$ from the anticentre ($\ell=180\arcdeg$), $\ell=130\arcdeg$
and $\ell=230\arcdeg$. These are shown as dashed lines in Figure~\ref{fig:arms}.
One might expect these sightlines to probe similar parts of the Galaxy. The angle between
the $\ell =230\arcdeg$ sightline and the normal to the
Perseus Arm is $\approx 45\arcdeg$, whereas the angle between the $\ell = 130\arcdeg$
sightline and the Perseus Arm normal is $\approx 30\arcdeg$. For the outer arm, the
offset between the $\ell = 130\arcdeg$ sightline and the normal to the arm is
$\approx 15\arcdeg$; the $\ell \approx 230\arcdeg$ sightline does not intersect any known
Outer Arm material. If the Outer Arm continued at the same pitch angle, the
$\ell = 230\arcdeg$ sightline would encounter the Outer Arm much further out,
$>20 \kpc$ from the Galactic Centre.

\begin{figure}
\includegraphics[width=0.5\textwidth]{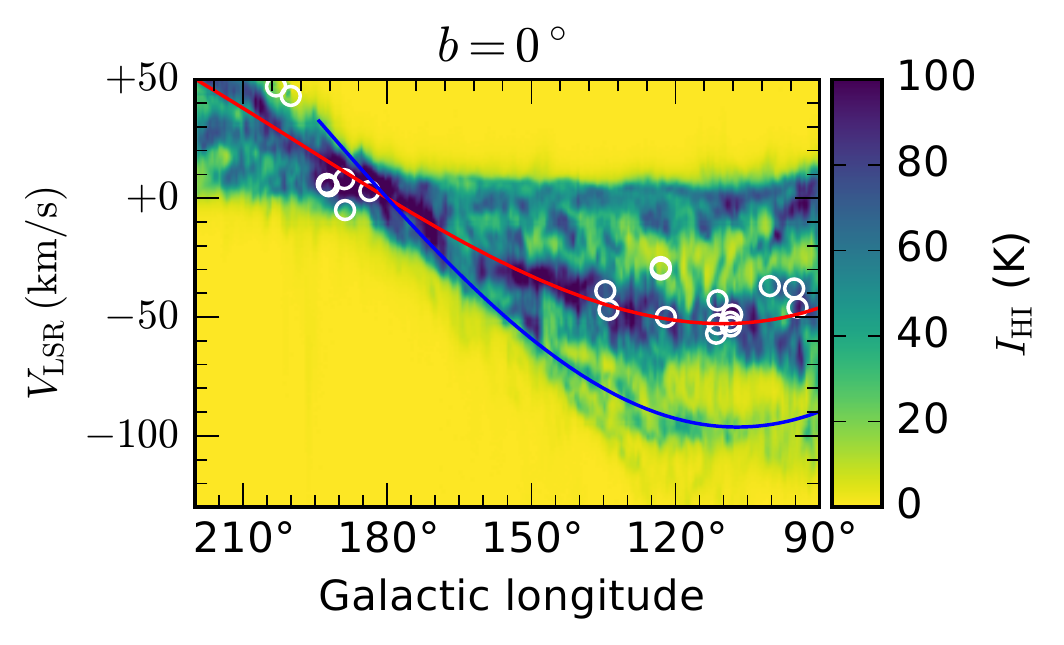}
\includegraphics[width=0.5\textwidth]{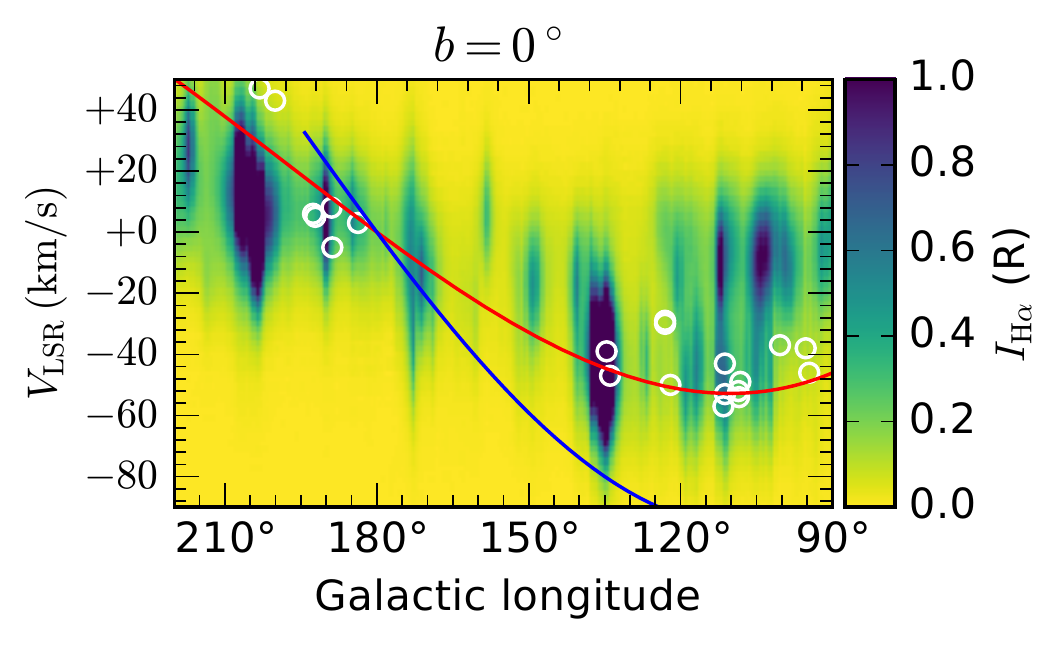}
\caption{Longitude-velocity diagrams of \Hi\ from LAB (top) and \Ha\ from WHAM (bottom). White circles show the longitudes and velocities of masers in the Perseus Arm from \citet{ReidMenten:2014}, as in Fig.~\ref{fig:arms}. Red and blue lines show the Perseus and Outer arms, respectively (see text).}
\label{fig:l-v}
\end{figure}

\begin{figure}
\includegraphics[width=0.5\textwidth]{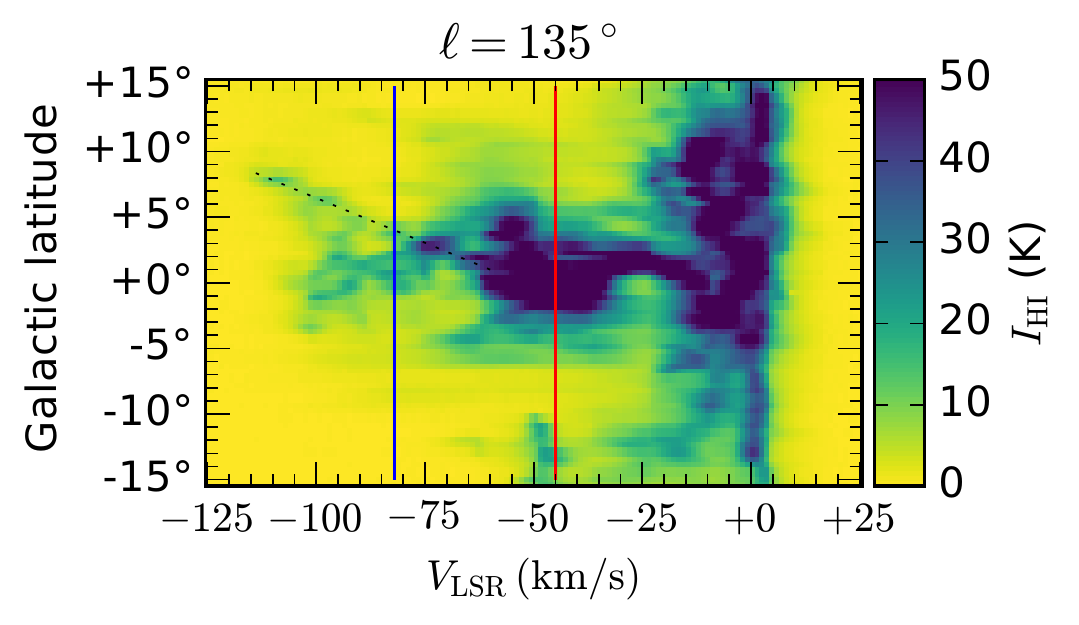}
\includegraphics[width=0.5\textwidth]{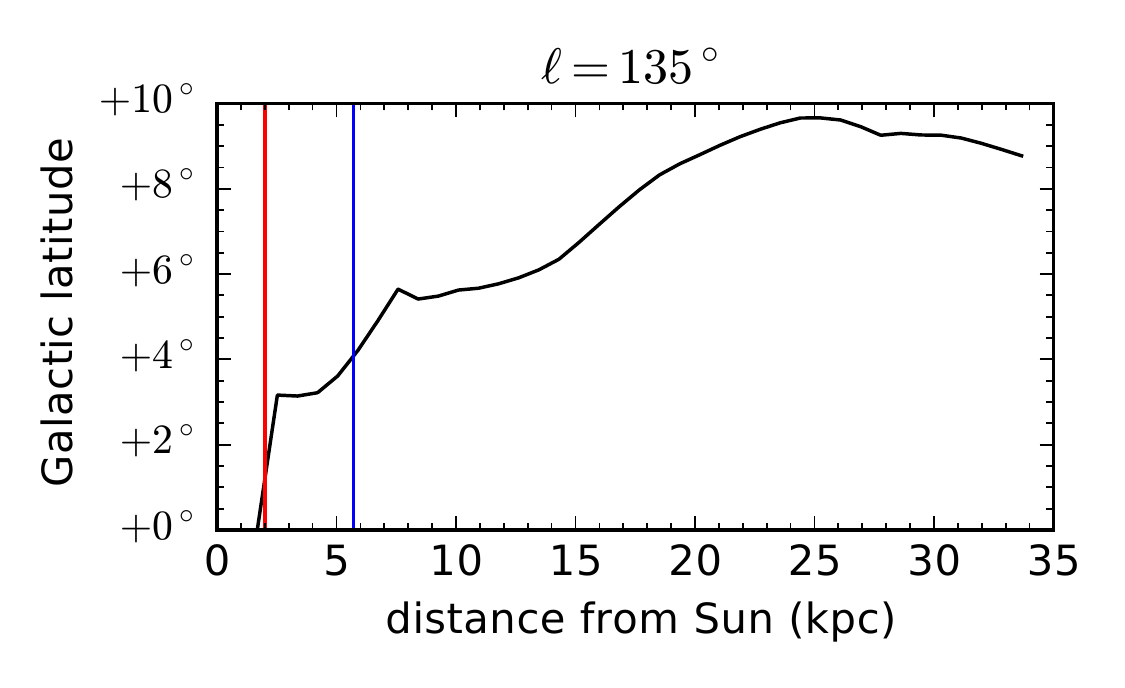}
\caption{{\em Top:} Latitude-velocity diagram of \Hi\ from LAB. The velocities
of the Perseus (red) and Outer (blue) arms at this longitude as shown in
Fig.~\ref{fig:l-v} are shown. A dotted line indicates the approximate position
of the warp emission.
{\em Bottom:} Latitude of the midplane of the warped Galaxy as a function of distance
from the Sun calculated using the fit to the LAB data
determined by \citet{KalberlaDedes:2007}. For a discussion of the considerable
uncertainties in this plot, see their paper. Distances to the arms at $\ell=135\arcdeg$
from the fit to maser distances \citep{ReidMenten:2014} are shown as an indication,
but the systematic uncertainties in the kinematic distances used to determine the
latitude of the warp are large.}
\label{fig:b-v}
\end{figure}

In Figure~\ref{fig:l-v}, we plot longitude-velocity diagrams of \Hi\ and \Ha\ to
describe the structure of Galactic emission in this region. The Perseus Arm is
the dominant feature, with the \Hi\ and \Ha\ emission brightest at $\vlsr
\approx -50 \kms$ in the midplane at $\ell = 135\arcdeg$. There are
kinematically-distinct \Hi\ components at local and Perseus Arm velocities.
Because the brightest \Ha\ emission is concentrated in \Hii\ regions, the \Ha\
is patchier than the \Hi. Typical \Ha\ line widths are $\approx 30 \kms$,
compared to $10 \kms$ for \Hi, which also makes it more difficult to separate
\Ha\ components. However, the local and Perseus Arm \Ha\ components are
kinematically distinct at $150\arcdeg \gtrsim \ell \gtrsim 130\arcdeg$
\citep{HaffnerReynolds:1999}, and Figure~\ref{fig:l-v} shows distinct components
at $\vlsr \approx -50 \kms$ and $\vlsr \approx -10 \kms$ extending to
$\ell \approx 100\arcdeg$. We adopt $-75 \kms < \vlsr < -30 \kms$ as the velocity
range which defines the Perseus Arm \Ha. 

The distance to masers in the Perseus Arm ranges from $1.9$ to $2.8 \kpc$ over
$120\arcdeg < \ell < 145\arcdeg$ with a line-of-sight depth of $0.38 \kpc$
from parallax measurements \citep[and references therein]{ReidMenten:2009,ReidMenten:2014}.
We show the longitudes and velocities of the masers \citeauthor{ReidMenten:2014} used to
determine this distance in Figure~\ref{fig:l-v}. The maser positions and
velocities trace both the \Ha\ and \Hi\ emission. We also plot lines showing the
longitude and velocity of the Perseus and Outer arms with velocities derived
using the linear function defining the Galactic rotation speed from
\citet[their Bayesian fit D1]{ReidMenten:2014}. Because kinematic distances in the
second quadrant overestimate the true source distance measured by parallax
\citep{FosterMacWilliams:2006,ReidMenten:2009} and the purpose of these lines is
simply to guide the eye, we multiplied the derived velocities for the arms by
$1.5$ and $1.3$, respectively, to match the lines to the kinematic features.

The latitude-velocity diagram in Figure~\ref{fig:b-v} shows the warp in the Fan
Region. There is evidence of a warp in the further reaches of the Perseus
Arm: the bright emission at $\vlsr \approx -45 \kms$ is centred at $b=0\arcdeg$ but gas
at more negative velocities predominantly lies above $b=0\arcdeg$.
At $\vlsr \approx -75 \kms$, the warp is evident at $b \approx +3\arcdeg$.
This is the Outer Arm, at a distance of $\approx 6 \kpc$
\citep{HachisukaBrunthaler:2009,ReidMenten:2014}. At $\vlsr \approx -100 \kms$,
the \Hi\ emission from the warp is brightest at $b \approx +6\arcdeg$, while it
extends to $b=+8\arcdeg$ at $\vlsr \approx -115 \kms$. In this region, there is
\Hi\ emission near $b=0\arcdeg$ out to large velocities in addition to the
warped emission.

In Figure~\ref{fig:b-v}, we also show a fit to the warp
along a sightline in the Fan Region, calculating following
\citet[with data provided by P.~M.~W. Kalberla, private communication, 2016]{KalberlaDedes:2007}.
In this fit, the emission is centred near $b=+3\arcdeg$ from
$1 \kpc \lesssim D \lesssim 5 \kpc$, near $b=+5.5\arcdeg$ from
$8 \kpc \lesssim D \lesssim 15 \kpc$, and near $b=+9\arcdeg$ over $D \gtrsim 20 \kpc$.
The emission at Perseus Arm velocities around $\ell=130\arcdeg$ is centred near
$b=+1\arcdeg$, while emission at Outer Arm velocities in this region is centred
near $b=+3 \arcdeg$ at a distance of $6 \kpc$
\citep{HachisukaBrunthaler:2009,ReidMenten:2014}.

There is no clear evidence of a
warp in \Ha\ emission, likely due to four factors: 1) the broader line widths of
\Ha\ ($\approx 30-40 \kms$, evident in Fig.~\ref{fig:l-v}) make it difficult to
separate the Outer Arm from the Perseus Arm in \Ha; 2) extinction makes distant
\Ha\ more difficult to observe \citep{MadsenReynolds:2005}; 3) the scale height
of the warm ionized medium (WIM) is large enough
\citep[$1-1.4 \kpc$;][]{HaffnerReynolds:1999,GaenslerMadsen:2008,SavageWakker:2009,Schnitzeler:2012}
that a $3\arcdeg$ offset of the midplane is a relatively small effect; and 4) the star
formation rate in the Outer Arm may be lower because star formation rates are typically
lower further from the Galactic Centre \citep{KennicuttEvans:2012}, producing a lower
ionizing flux and thus less \Ha\ emission.

\section{Results} 
\label{sec:results}

\begin{figure*}
\includegraphics[width=0.45\textwidth]{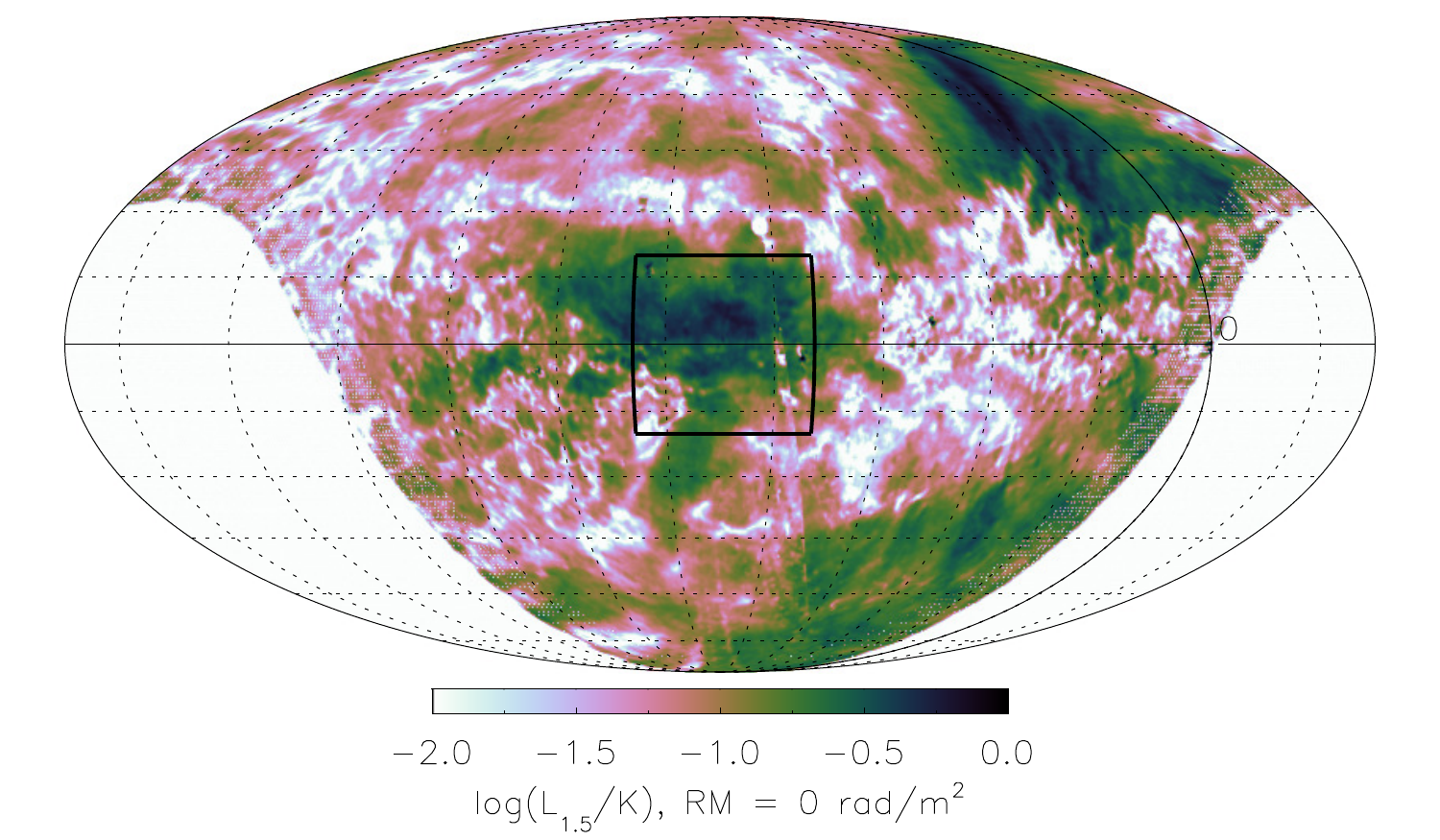}
\includegraphics[width=0.45\textwidth]{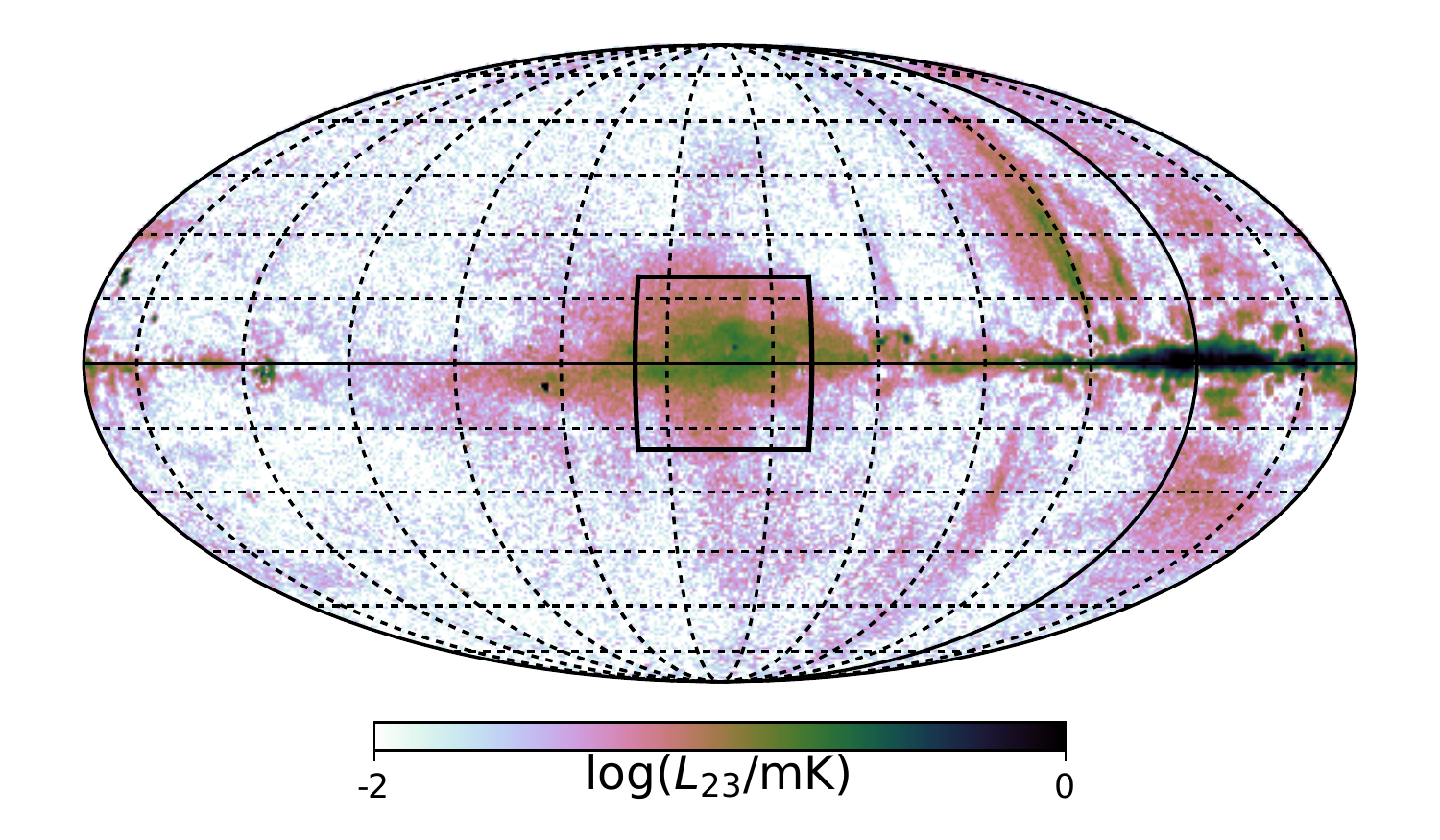} \\
\includegraphics[width=0.45\textwidth]{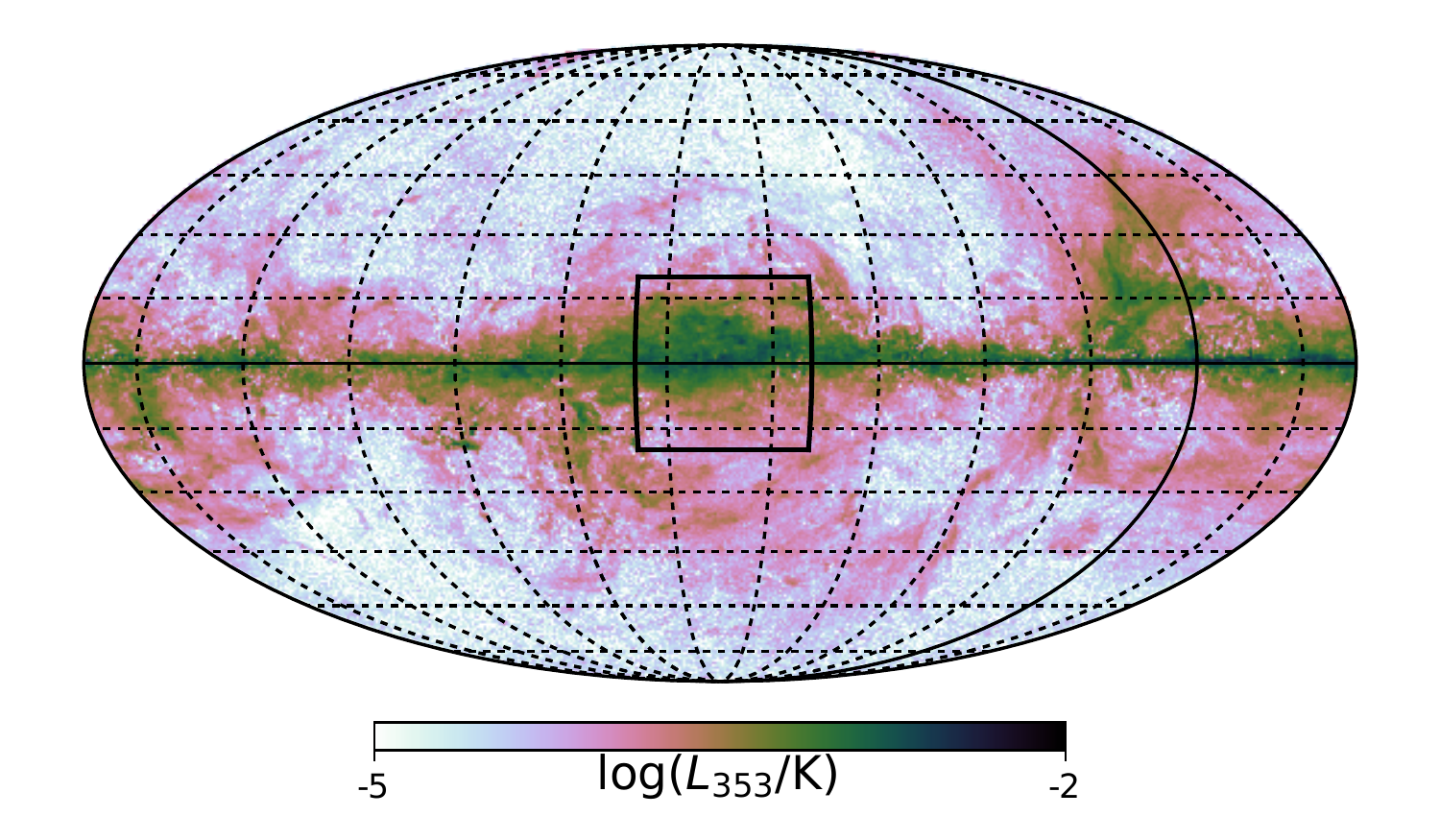}
\includegraphics[width=0.45\textwidth]{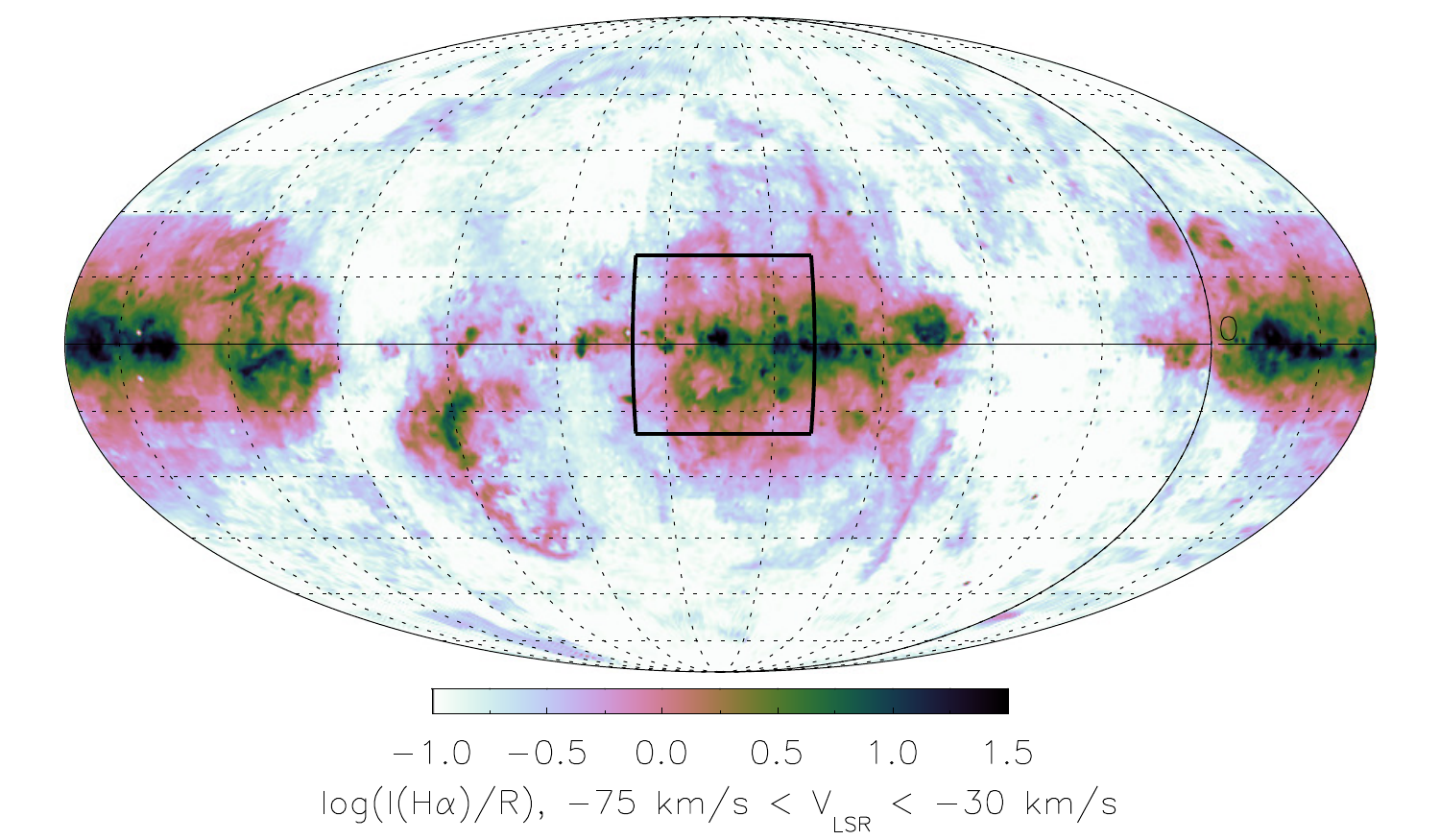}
\caption{All-sky images centred on $\ell=135\arcdeg$ and $b=0\arcdeg$ in a Mollweide
projection with grid lines every $\Delta \ell = 30\arcdeg$, $\Delta b = 15\arcdeg$.
Solid lines show $\ell = 0\arcdeg$ and $b = 0\arcdeg$.
The intensity scales in each image are logarithmic.
Top left: GMIMS-HBN $1.5 \GHz$ polarized intensity at Faraday depth $0 \radmsq$.
Top right: WMAP $23 \GHz$ linearly polarized intensity \citep{BennettLarson:2013}.
Bottom left: Planck $353 \GHz$ linearly polarized intensity from the 2015 data release
\citep{Planck-CollaborationAdam:2016} smoothed with a $20'$ FWHM Gaussian beam.
Bottom right: WHAM-SS \Ha\ emission \citep{HaffnerReynolds:2003,HaffnerReynolds:2010} integrated over
$-75 \kms < \vlsr < -30 \kms$, isolating emission from the Perseus Arm and beyond.
Boxes indicate the corners of the region shown in Fig.~\ref{fig:multifreq}.}
\label{fig:allsky}
\end{figure*}

\subsection{Morphology of polarized emission} \label{sec:morphology}

In Figure~\ref{fig:allsky} we show the Fan Region in its Galactic context. The
all-sky images are centred on the second quadrant
($90\arcdeg \lesssim \ell \lesssim 180\arcdeg$), where the Fan Region is located.
The Fan Region is identified by the bright signal in polarized intensity at
$1.5 \GHz$ ($L_{1.5}$), $23 \GHz$ ($L_{23}$), and $353 \GHz$ ($L_{353}$).
At $1.5 \GHz$, the polarized emission is centred above the plane, with
$L_{1.5} \gtrsim 0.1 \K$ (green in Fig.~\ref{fig:allsky}) emission extending over
$-12\arcdeg \lesssim b \lesssim +25\arcdeg$, $90\arcdeg \lesssim \ell \lesssim 180\arcdeg$.
The $23 \GHz$ and $353 \GHz$ emission are roughly similar, in both cases with
emission centred at $b \approx +4\arcdeg$.

\begin{figure*}
\includegraphics[width=0.45\textwidth]{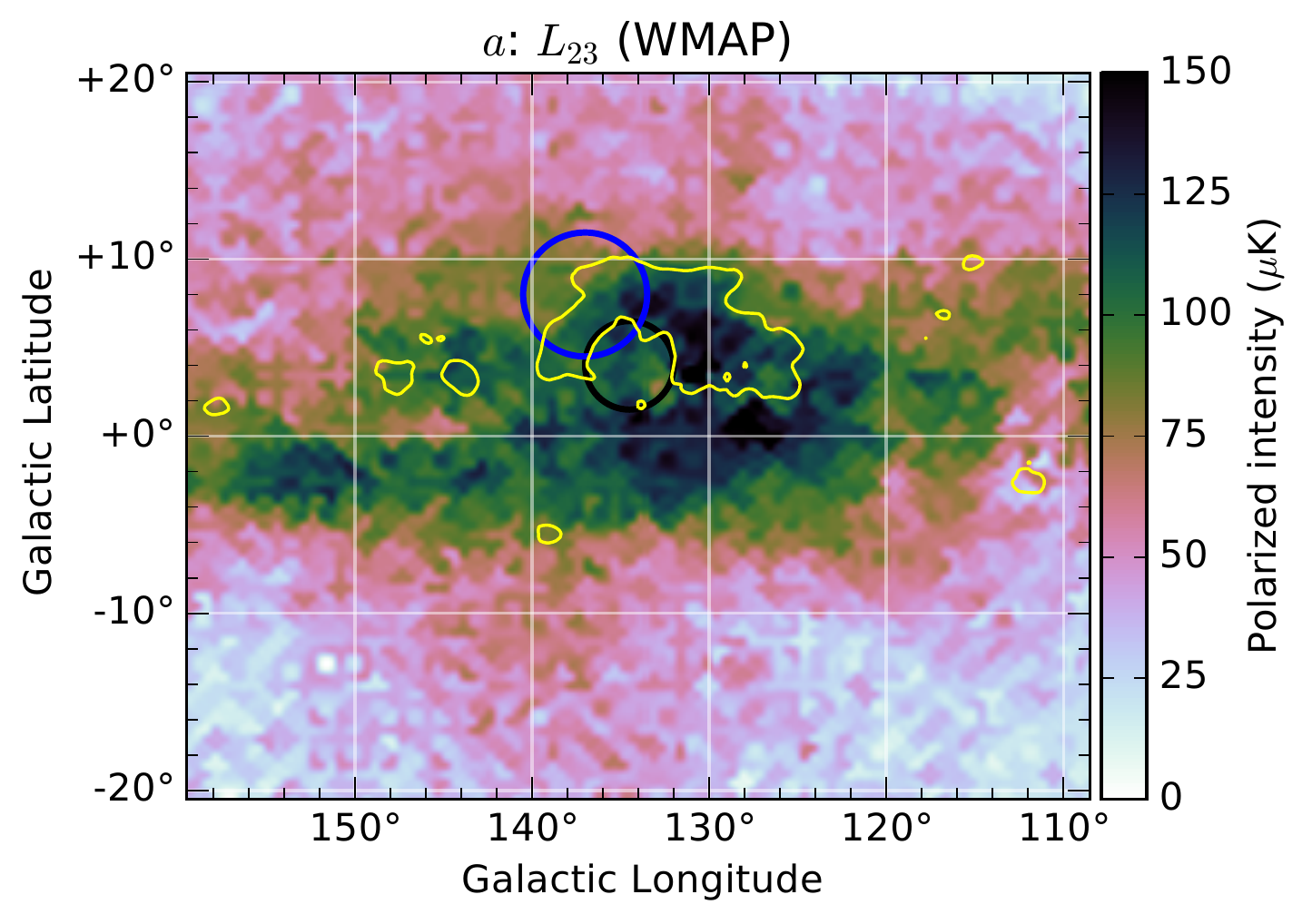}
\includegraphics[width=0.45\textwidth]{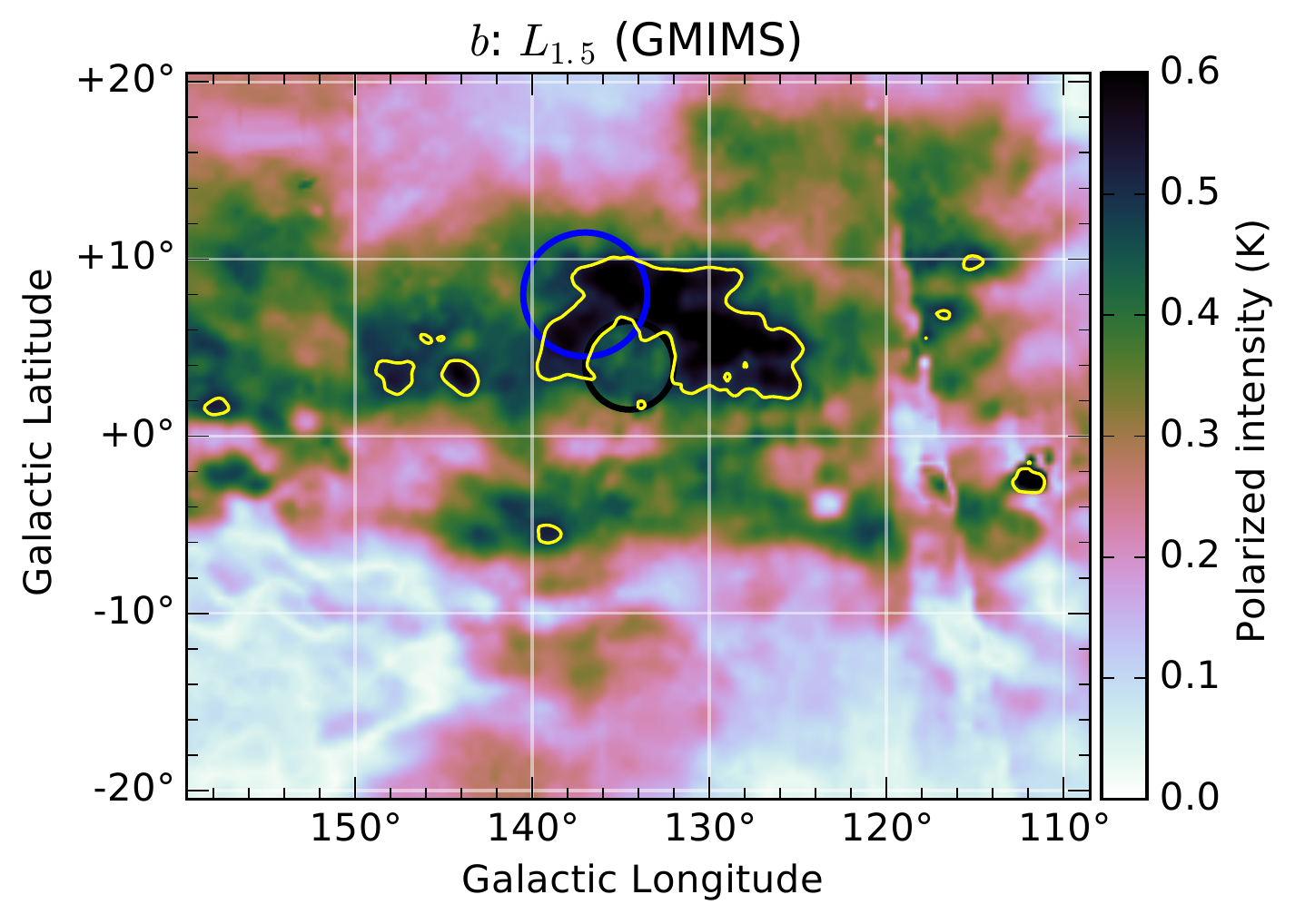} \\
\includegraphics[width=0.45\textwidth]{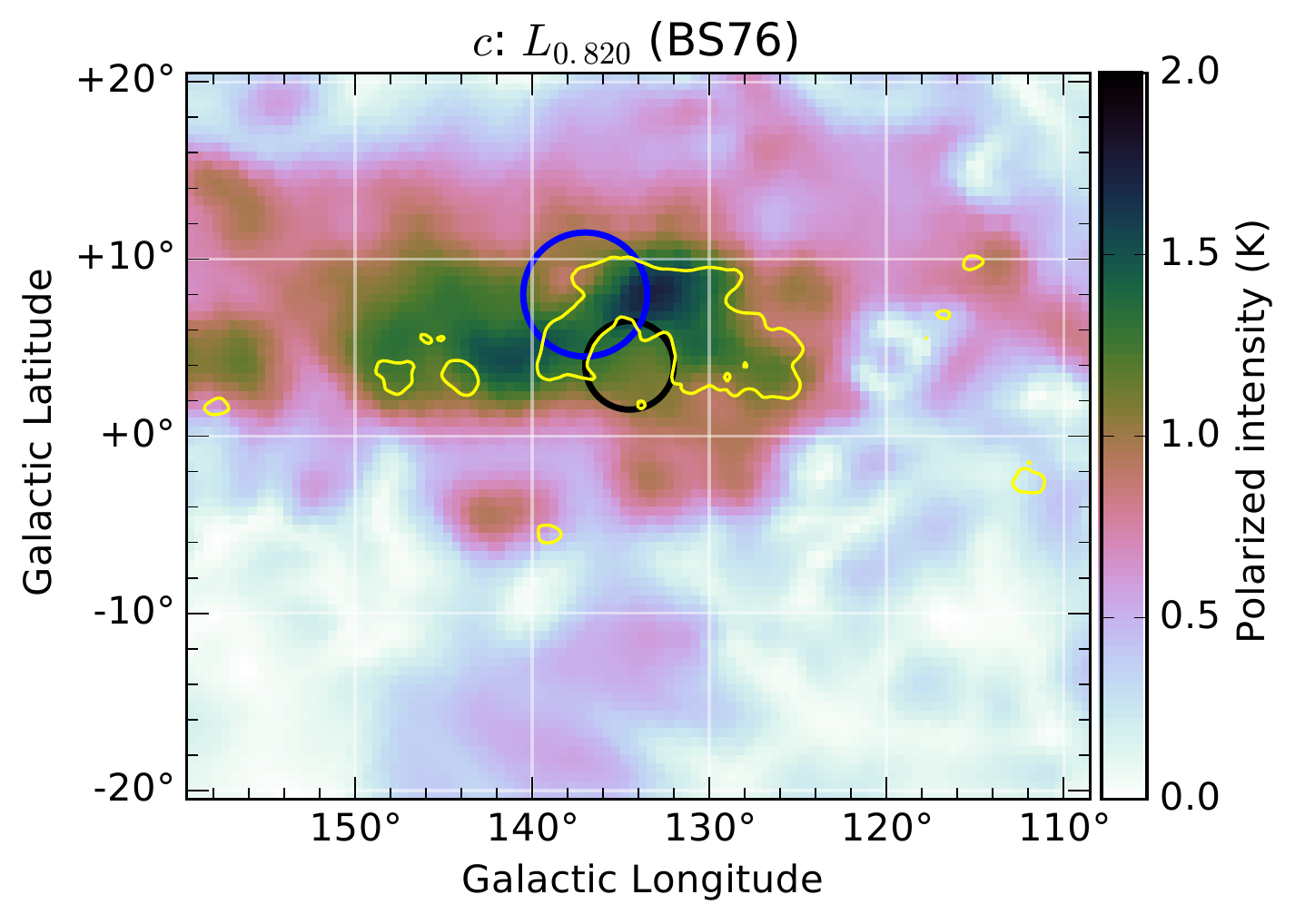}
\includegraphics[width=0.45\textwidth]{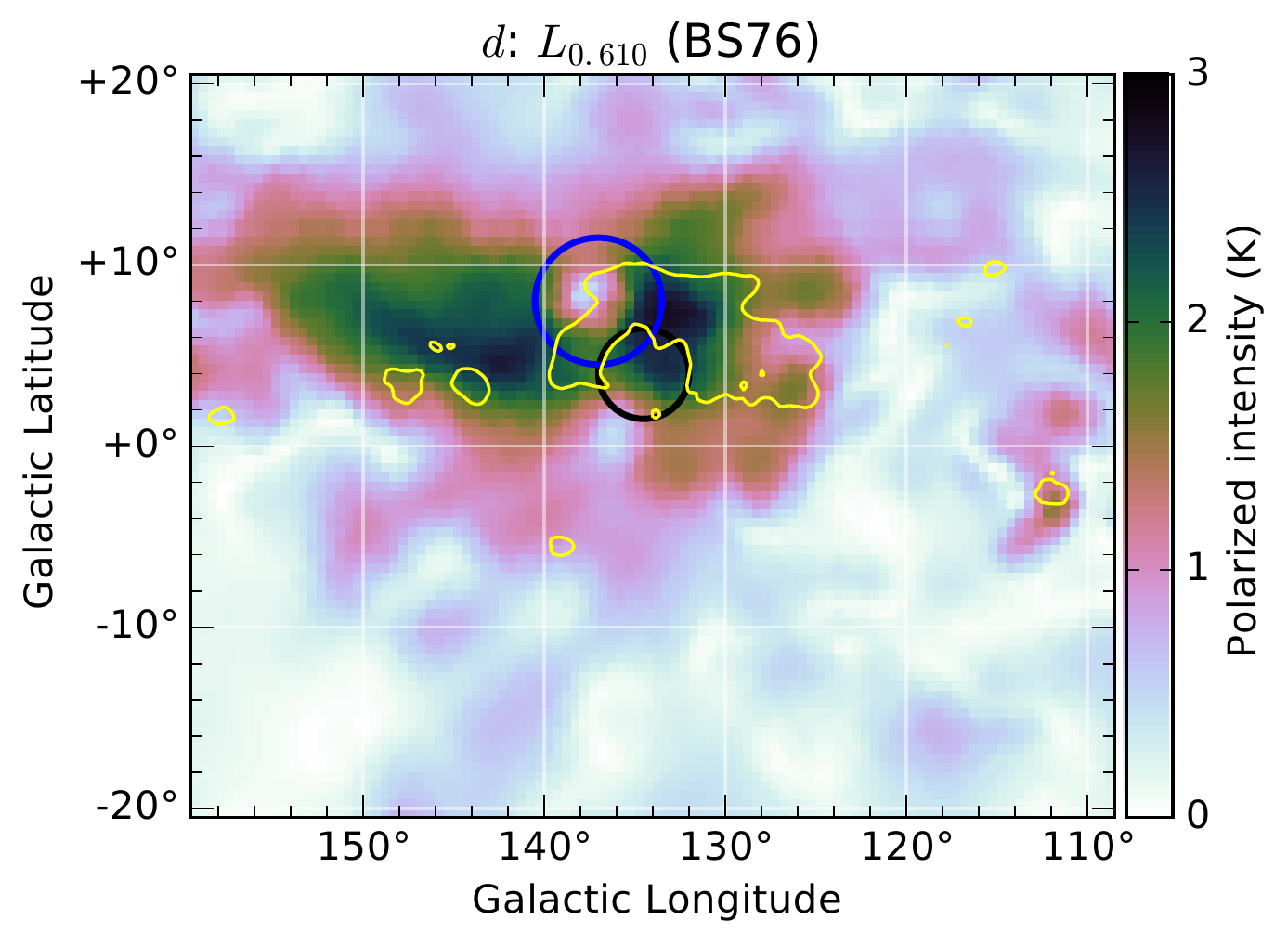} \\
\includegraphics[width=0.45\textwidth]{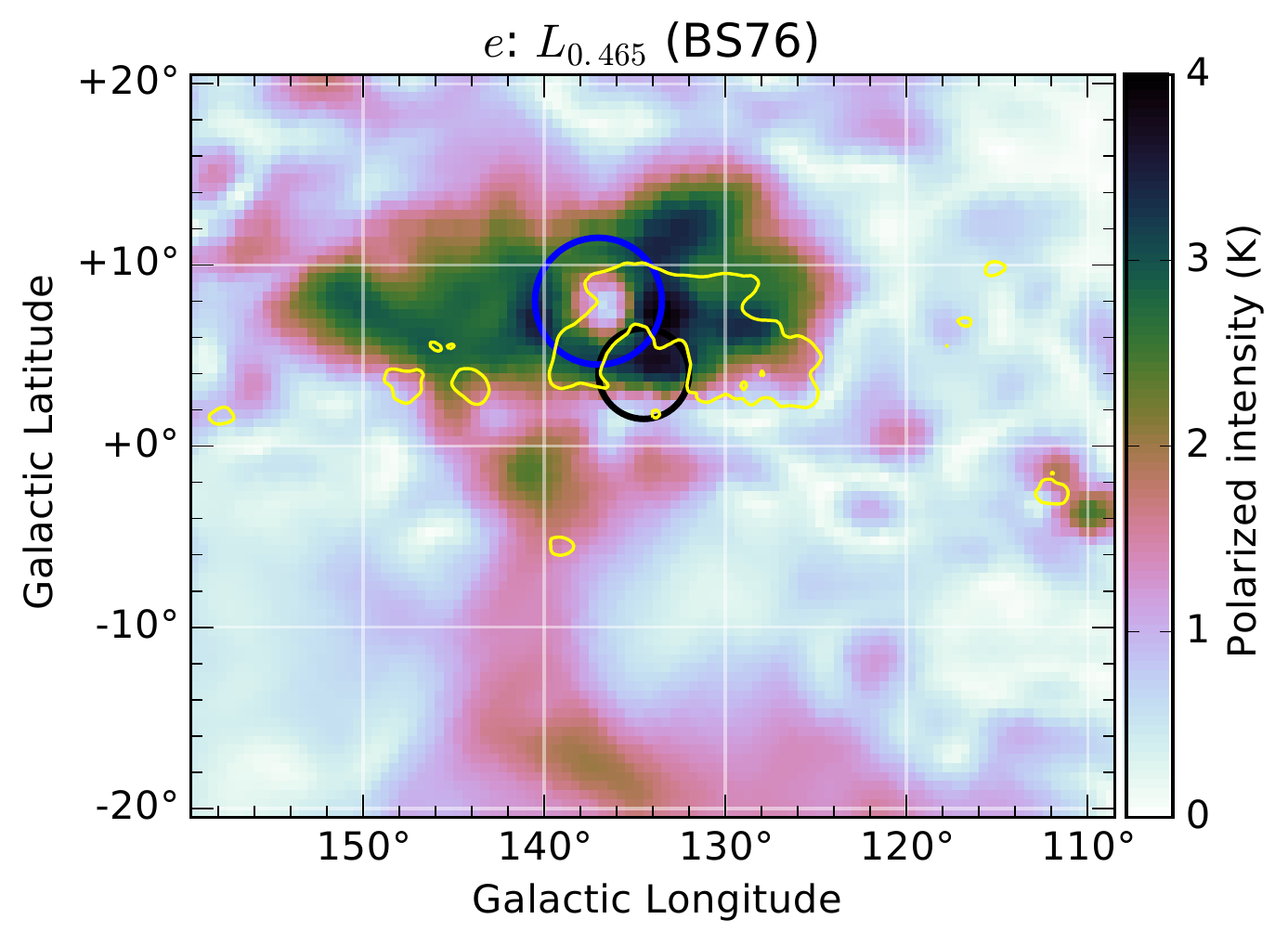}
\includegraphics[width=0.45\textwidth]{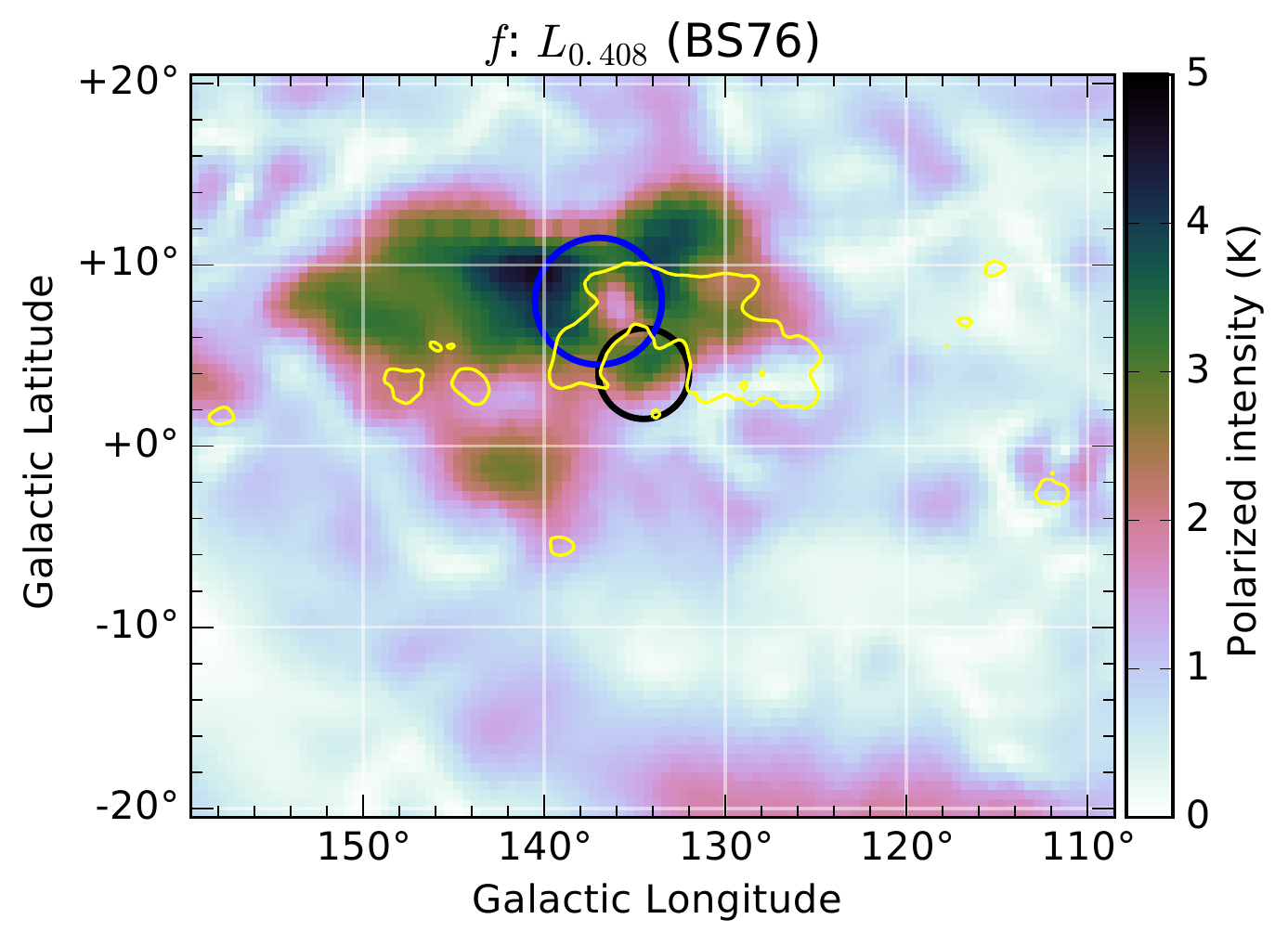}
\caption{Images of polarized intensity at $23 \GHz$ (WMAP), $1.5 \GHz$ (GMIMS), and $408-820 \MHz$ \citep{BrouwSpoelstra:1976,CarrettiBernardi:2005}.
Yellow contours show GMIMS data ($L_{1.5}$) at an antenna temperature of $0.5 \K$.
Black and blue circles show the features at $(134.5\arcdeg, +4\arcdeg)$ and
$(137\arcdeg, +8\arcdeg)$ discussed in the text.
A small artefact remains in the GMIMS $L_{1.5}$ image at 0 hours right ascension
(the diagonal stripe around $\ell = 115\arcdeg$ in panel $b$).}
\label{fig:multifreq}
\end{figure*}

\begin{figure}
\includegraphics[width=0.5\textwidth]{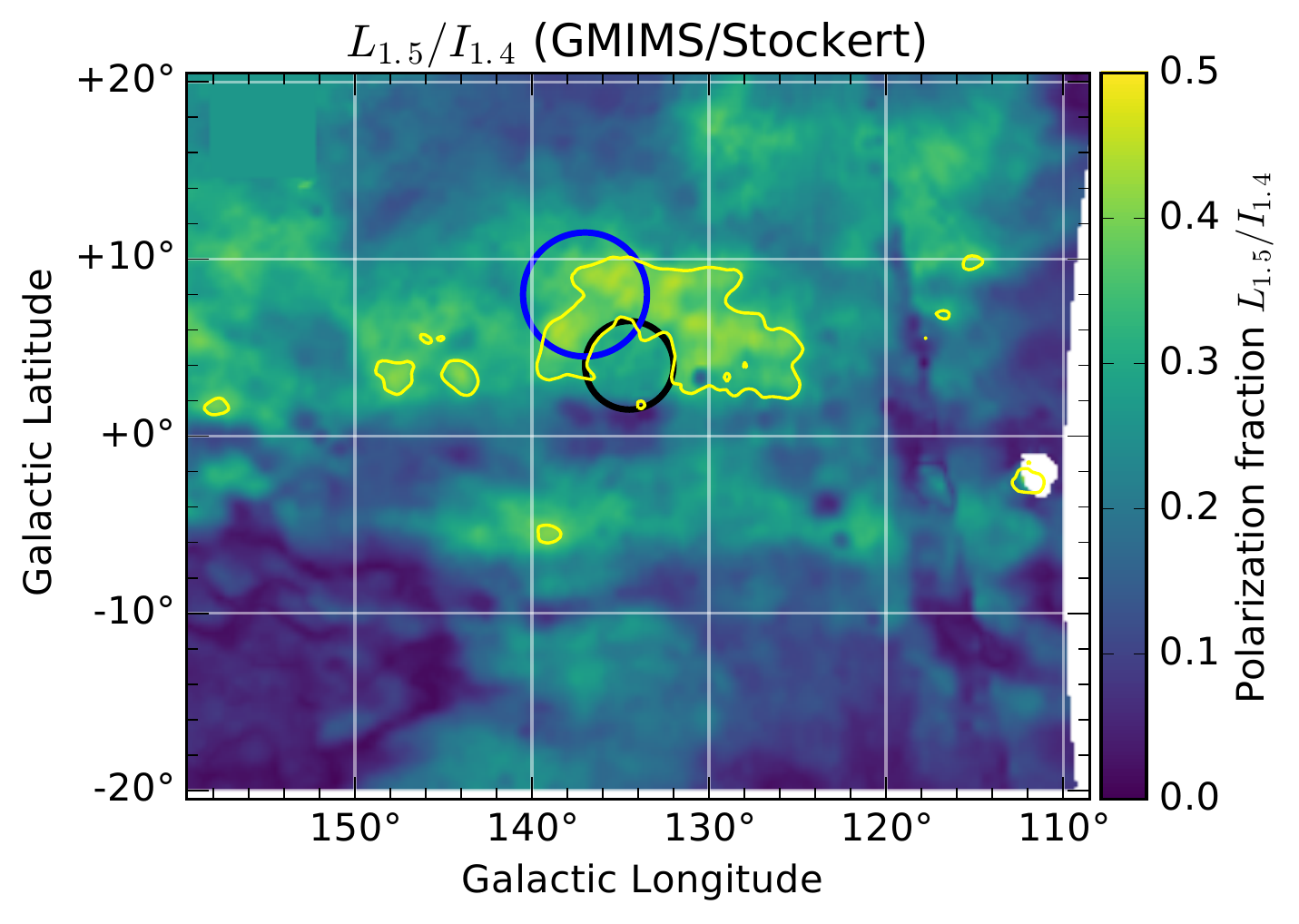}
\caption{Polarization fraction from comparing the GMIMS data to the Stockert
$1.4 \GHz$ Stokes $I$ data \citep{Reich:1982va}. Yellow contours show
$L_{1.5} = 0.5 \K$, as in Fig.~\ref{fig:multifreq}.}
\label{fig:polfrac}
\end{figure}

The brightest emission, $L_{1.5} \gtrsim 0.3 \K$, is mostly in a smaller region,
$120\arcdeg < \ell < 150\arcdeg$ and $-6\arcdeg < b < +12 \arcdeg$.
We show maps of polarized intensity in this region from $408 \MHz - 23 \GHz$
in Figure~\ref{fig:multifreq}. 
The morphology is qualitatively
different at $\nu \lesssim 600 \MHz$ and $\nu \gtrsim 1 \GHz$. At high
frequencies, the emission fills most of this region. At $23 \GHz$, the emission
is bright both at $|b| \lesssim 2\arcdeg$ and above the plane up to $+10\arcdeg$. In
contrast, at $1.5 \GHz$ and all lower frequencies, the emission is much fainter
in the plane ($|b| \lesssim 2\arcdeg$) than at $b \approx +8\arcdeg$. However, the
bright emission is similar in morphology at $23$ and $1.5 \GHz$ at $b \gtrsim +3\arcdeg$.
At both frequencies, the bright emission follows an arc from around $(134\arcdeg,
+8\arcdeg)$ to $(128\arcdeg, +3\arcdeg)$, emphasized by the yellow contours in
Figure~\ref{fig:multifreq}.

There is no obvious feature corresponding to the Fan Region in total intensity
at $1.4 \GHz$.
We show the polarization fraction at $\sim 1.5 \GHz$ in Figure~\ref{fig:polfrac}. The
structure of the polarization fraction image closely traces that of the polarized
intensity image (Fig.~\ref{fig:multifreq}$b$) because the Stokes $I$ emission
is much more uniform than the polarized intensity. The polarization fraction is
highest, $\approx 40\%$, in the region where $L_{1.5}$ is brightest, in the arc
which peaks at $(134\arcdeg, +8\arcdeg$).

There is a patch where the polarized intensity at $1.5 \GHz$ is lower, centred at $(134.5\arcdeg, +4\arcdeg)$ and $\approx 5\arcdeg$ in diameter; this area, shown with black circles in Figures~\ref{fig:multifreq}--\ref{fig:maps} as well as Figures \ref{fig:cgps}--\ref{fig:maps6cm} below, is a major focus of this paper. The reduced $L_{1.5}$ cannot be entirely due to Faraday depolarization because it is also seen to some extent in $L_{23}$. However, the $1.5 \GHz$ fractional polarization in Figure~\ref{fig:polfrac} is lower in that patch ($\approx 30\%$) than in the surrounding Fan Region ($\approx 40\%$).

This reduction in polarized intensity is not seen in the Dwingeloo 1411~MHz data \citep{BrouwSpoelstra:1976,CarrettiBernardi:2005}. That survey is sparsely sampled, and by checking the measured data points we have verified that there was no measurement near $(134.5\arcdeg, +4\arcdeg)$: the 1411~MHz survey could not have detected this feature. A similar check of the 820~MHz observation points shows that there was a measurement close to this position, and the interpolated image at 820~MHz does indeed show a decline in polarized intensity at this position (Figure~\ref{fig:multifreq}$c$).

Somewhat puzzling is the fact that the depression in polarized intensity is not seen in the \citet{WollebenLandecker:2006} data. That survey was made using drift scans, and a check showed that drift scans were made across this region at full sampling; however, no cross scans in declination were made. We regard the GMIMS data as more reliable. Scanning was in two directions, fully Nyquist sampled, and the area of interest was crossed by many scans. The intersecting scans were reconciled by the basketweaving technique. Inspection of the GMIMS datacube shows that the depression in polarized intensity is seen at every frequency across the GMIMS band as well as in the image after RM synthesis.

At the lowest frequencies, the polarized emission is brightest in a ring centred
at $(137\arcdeg, +8\arcdeg)$, shown with blue circles in Figure~\ref{fig:multifreq}.
At $150-350 \MHz$, this ring is the dominant feature
identifiable in the Fan Region, with a diameter of about $7\arcdeg$ in Westerbork Synthesis Radio Telescope data.
Various authors have used Westerbork data to describe the ring as a relic Str\"omgren
sphere at a distance of $\approx 200 \pc$ \citep{IacobelliHaverkorn:2013a} and
as a depolarization artefact of a uniform Faraday depth distribution
\citep{HaverkornKatgert:2003a}. The Westerbork data use aperture
synthesis and thus are not sensitive to features $\gtrsim 10\arcdeg$ in size
\citep{BernardiBruyn:2009}, whereas the \citet{BrouwSpoelstra:1976} data are
single-antenna and thus should be sensitive to emission on all resolved scales.
The ring is also the dominant feature
at frequencies up to $610 \MHz$ (see Figure~\ref{fig:multifreq}).
Its diameter is
larger at higher frequencies, and it also becomes less clearly defined as a
circular feature. At $\nu \gtrsim 1.5 \GHz$, the ring is not apparent. At $820
\MHz$, both the ring and the broader, high-frequency feature are evident.

\subsection{Morphological comparison of \Ha\ and $\mathbf{1-23} \GHz$ polarized emission}
\label{sec:ha_comp}

\begin{figure*}
\includegraphics[width=0.9\textwidth]{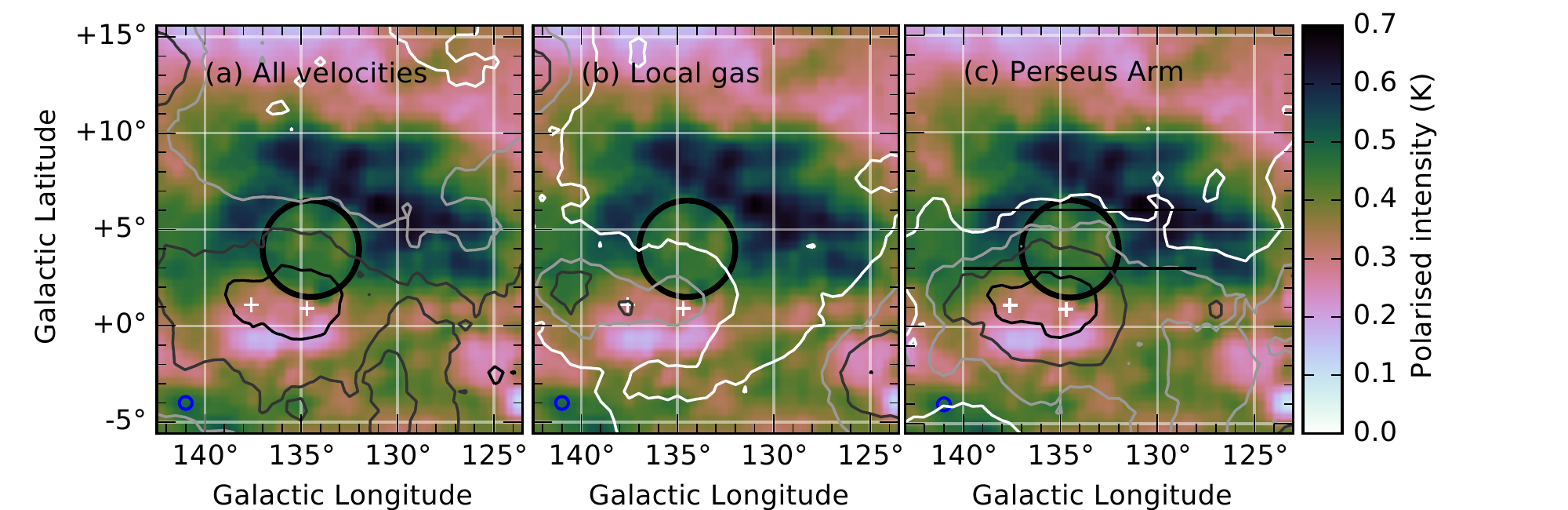}
\includegraphics[width=0.9\textwidth]{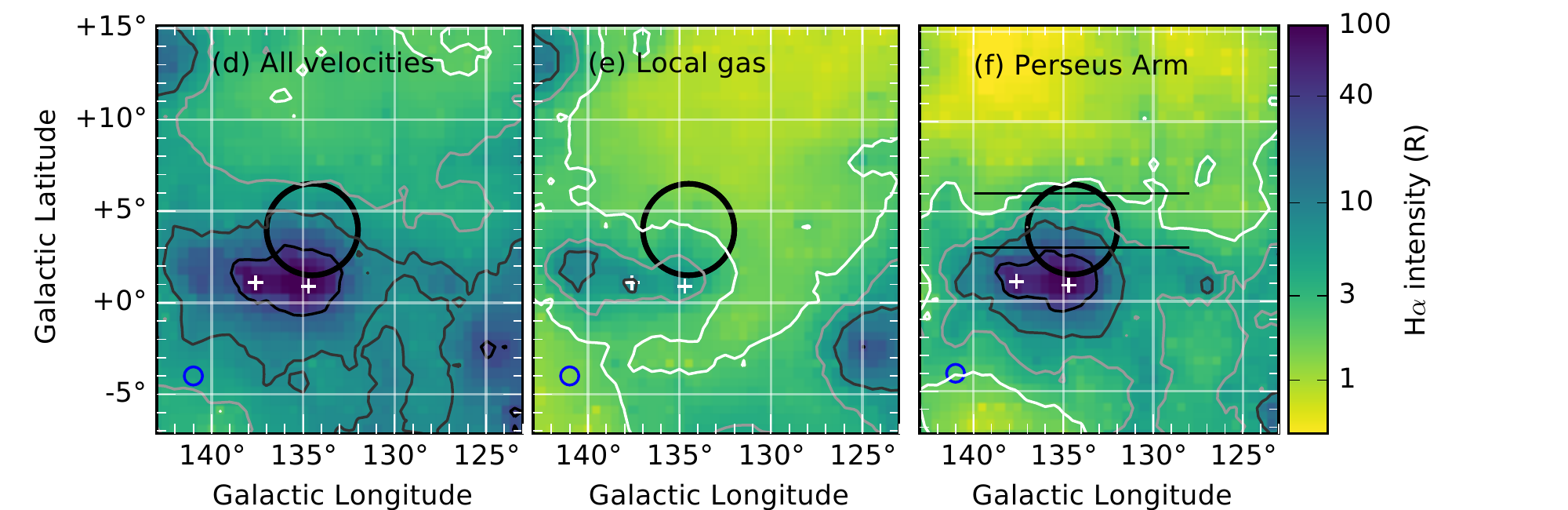}
\caption{Top row: $1.5 \GHz$ polarized continuum emission from GMIMS (as in
Fig.~\ref{fig:multifreq}$b$); we show the same map in all three columns.
Bottom row: WHAM-SS \Ha\ maps in three velocity ranges. Left: all velocities
with significant emission ($-75 \kms < \vlsr < +15 \kms$).
Middle: local velocities ($ |\vlsr| < 15 \kms$). Right: Perseus Arm velocities
($-75 \kms < \vlsr < -30 \kms$). Corresponding \Ha\ contours at intensities of
$2$ (white), $4$ (light gray), $8$ (dark gray), and $32 \R$ (black) are shown in both
rows. White $+$ signs show the centres of W4 and W5. Horizontal lines in panels $c$ and $f$ show
the region used to calculate Fig.~\ref{fig:depol_ha}. Black circles are as in
Fig.~\ref{fig:multifreq}. $1.5 \GHz$ data are shown with a linear
scale; \Ha\ data are shown with a logarithmic scale. The GMIMS (top) and WHAM
(bottom) beams are shown with blue circles in the lower-left corner.}
\label{fig:maps}
\end{figure*}

The lower-right panel of Figure~\ref{fig:allsky} shows \Ha\ emission with a velocity
criterion ($-75 \kms < \vlsr < -30 \kms$) which excludes local gas but includes
emission from the Perseus Arm and more distant gas
\citep{{HaffnerReynolds:1999,MadsenReynolds:2006}}. We compare $L_{1.5}$ with \Ha\
at the Perseus Arm velocity
as well as at local velocities and integrated over all velocities in Figure~\ref{fig:maps}.
As in the polarized continuum maps, the \Ha\ emission is brightest around
$100\arcdeg \lesssim \ell \lesssim 150\arcdeg$, although the \Ha\ emission is
centred at somewhat lower latitudes ($b \approx 0\arcdeg$) and longitudes than
the polarized continuum. The highest $1.5 \GHz$ polarized intensity,
$L_{1.5} \approx 0.65 \K$, is in a region (near $(134\arcdeg, +8\arcdeg)$) with
little \Ha\ emission at any velocity.
The region with depressed $L_{1.5}$ around $(134.5\arcdeg, +4\arcdeg)$ (which we
introduced in Section~\ref{sec:morphology}) corresponds with increased \Ha\
intensity at Perseus Arm velocities. The white ($I_{\Ha} = 2.0 \R$\footnote{$1
\textrm{ Rayleigh (R)}= 10^6 / (4 \pi) \textrm{ photons cm}^{-2} \textrm{ s}^{-1} \textrm{ sr}^{-1}$.})
contour in Fig.~\ref{fig:maps}$c$ traces the lower envelope of the bright
($L_{1.5} \gtrsim 0.5 \K$; dark blue in Fig.~\ref{fig:maps}$a$--$c$) polarized
emission, suggesting qualitatively that the polarized intensity in this region is
anti-correlated with the Perseus Arm \Ha\ intensity.

\begin{figure}
\includegraphics[width=0.5\textwidth]{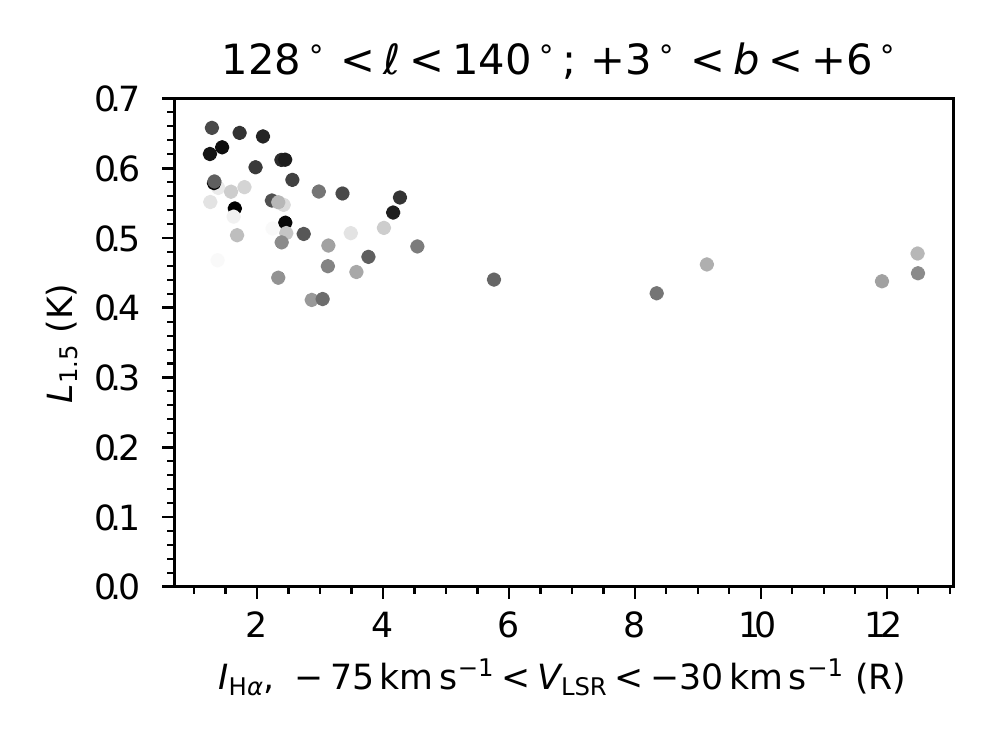}
\caption{Polarized intensity ($L_{1.5}$) as a function of integrated \Ha\ intensity from the Perseus Arm ($-75 \kms < \vlsr < -30 \kms$). Each point denotes $I_{\Ha}$ from a single WHAM beam compared to the average GMIMS $L_{1.5}$ within the beam in between the lines shown in Fig.~\ref{fig:maps}. Points are colour-coded for Galactic longitude, with the lightest points at $\ell=140\arcdeg$ and black points at $\ell = 128\arcdeg$.}
\label{fig:depol_ha}
\end{figure}

\label{sec:depol}

We quantify the relationship between $L_{1.5}$ and $I_{\Ha}$ from the Perseus
Arm in the scatter plot in Figure~\ref{fig:depol_ha}. We chose a narrow latitude range
to minimize the impact of the latitude dependence of $I_{\Ha}$ on the scatter plot.
The latitude range $+3\arcdeg < b < +6\arcdeg$ cuts through the region of reduced
$L_{1.5}$ around $(134.5\arcdeg, +4\arcdeg)$ we discussed in Section~\ref{sec:morphology}.
We chose the longitude range $128\arcdeg < \ell < 140\arcdeg$ to encompass the bright
emission from the Fan Region as seen at $1.5 \GHz$, $23 \GHz$, and $353 \GHz$
(see Figs.~\ref{fig:allsky} and \ref{fig:multifreq}).
In the six sightlines in the plot with $I_{\Ha}>5 \R$,
the polarized intensity is very consistent, $\langle L_{1.5} \rangle = 0.44 \K$
with standard deviation $\sigma_{L1.5} = 0.02 \K$. In sightlines with
$I_{\Ha} < 3 \R$, there is a wider range of polarized intensities, typically
$\approx 30\%$ higher than in the \Ha-bright sightlines.
The anticorrelation of $L_{1.5}$ with $I_{\Ha}$ is most likely due to depolarization,
not an intrinsic absence of polarized emission along these sightlines. We expect
depolarization due to
all gas in front of the emission, so correlating structures in depolarization
with \Ha\ emission at an estimated distance measures the minimum distance to the
emitting gas. The \Ha\ emission from the Perseus Arm (Fig.~\ref{fig:maps}$c$ and $f$) is
brightest in a roughly circular region centred on the W3/W4/W5 complex of \Hii\
regions. (W4 is shown with a $+$ sign in Fig.~\ref{fig:maps}.) From the optical
line ratios, this bright
\Ha\ emission is more typical of O~star \Hii\ regions than of the diffuse WIM
\citep{MadsenReynolds:2006}, indicating that the gas is primarily photoionized
by O~stars in those \Hii\ regions. The arched shape of the $1.5 \GHz$
emission is matched by an arch in the Perseus Arm \Ha\ contour at ($134\arcdeg$,
$+7\arcdeg$) in Figure~\ref{fig:maps}$c$. This fainter \Ha\ from the Perseus Arm,
which is spatially coincident with the high polarized intensity from the Fan Region,
has optical line ratios (especially $\nii \lambda 6584/\Ha$) which are higher and thus more
WIM-like than the \Hii\ region emission \citep[see Figure~11 of][]{MadsenReynolds:2006}.

We have not applied an extinction correction to the \Ha\ data. From
a comparison of \Ha\ and \Hb\ observations, there are $A_V = 3.13 \pm 0.2$ magnitudes of
extinction towards the W4 \Hii\ region at \Ha\ at ($134.7\arcdeg$, $+0.9\arcdeg$),
the brightest sightline in this region in \Ha\ \citep{MadsenReynolds:2006}.
Robust extinction measurements are only available for a small number of individual
sightlines, so a systematic extinction correction is difficult. We expect that
the dust column and thus the extinction is highest in sightlines which are brightest
in \Ha, so we expect that the main impact of an extinction correction would be
to increase the contrast between the faint sightlines and the bright ones. This
would not change our interpretation of Figures~\ref{fig:maps} and \ref{fig:depol_ha}:
the sightlines with high $L_{1.5}$ would still be the sightlines with the lowest
$I_{\Ha}$.

\begin{figure}
\includegraphics[width=0.5\textwidth]{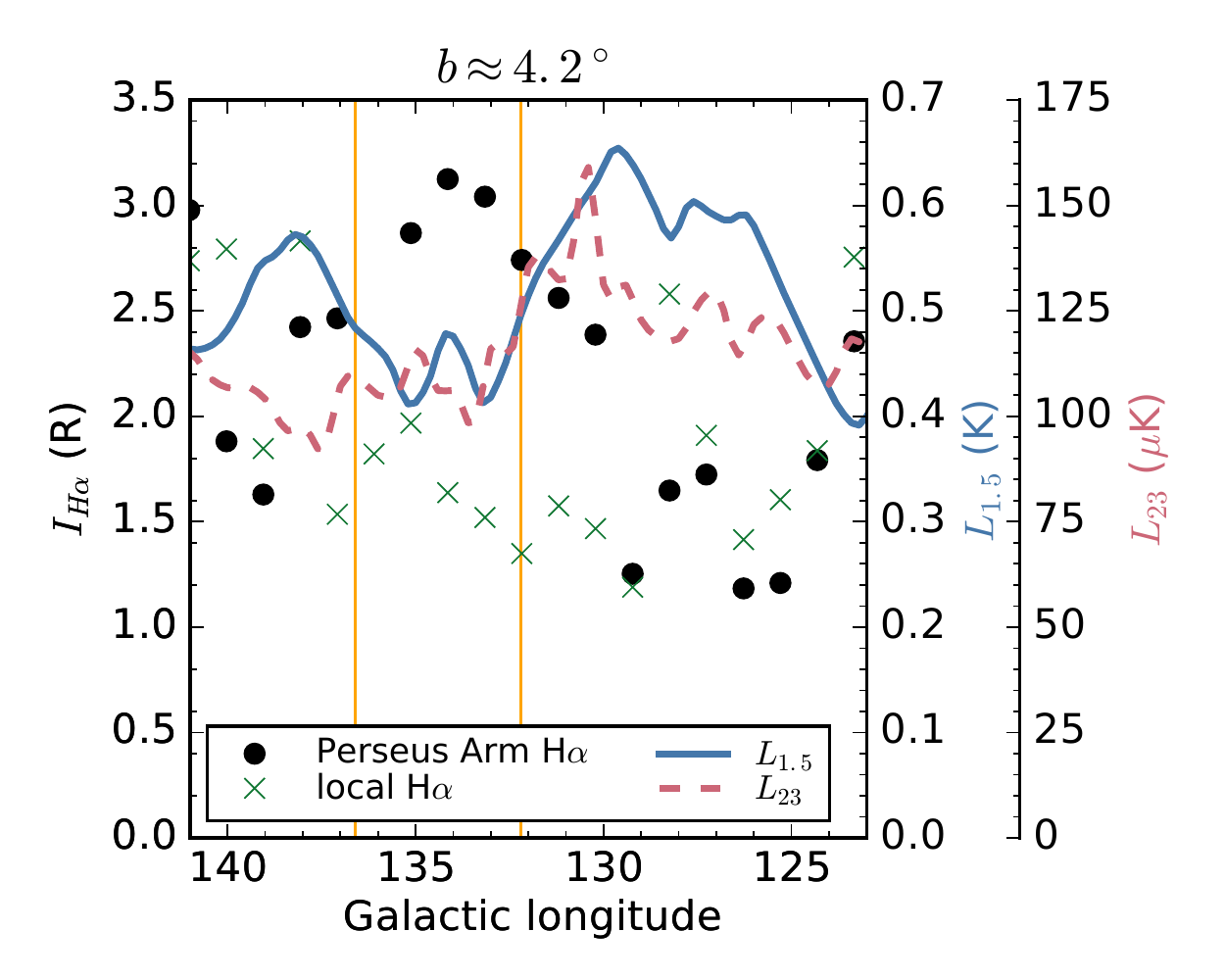}
\caption{Polarized intensity ($L_{1.5}$ and $L_{23}$), \Ha\ intensity at Perseus Arm velocities ($-75 \kms < \vlsr < -30 \kms$), and \Ha\ intensity at local velocities ($|\vlsr| < 15 \kms$) as a function of Galactic longitude in the range $+4\arcdeg \le b \le +5\arcdeg$. The \Ha\ points shown are from individual WHAM beams at $b=+4.2\arcdeg$ because the WHAM survey is not Nyquist sampled. Vertical orange lines show the outer boundaries of the W4 superbubble walls from $I_{4.8}$ data (see Fig.~\ref{fig:maps6cm} below).}
\label{fig:depol_long}
\end{figure}

In Figure~\ref{fig:depol_long}, we show a slice of $1.5$ and $23 \GHz$ polarized
intensities and the Perseus Arm \Ha\ intensity. In this slice, the \Ha\ intensity
peaks at $\ell=134\arcdeg$, the location of the minimum in $L_{1.5}$. At $1.5 \GHz$,
the polarized intensity peaks at $\ell \approx 130\arcdeg$ and $138\arcdeg$, areas of
relatively low \Ha\ intensity, and has a local minimum near $\ell=134\arcdeg$.
A similar trend is evident at $820 \MHz$ (not shown) but not at lower radio
frequencies. This trend is also evident at $23 \GHz$, although the local maximum at
$\ell = 138\arcdeg$ is not present. The figure also shows the local
\Ha\ emission, which does not have a discernible peak either correlated or
anti-correlated with the polarized intensity. This indicates that the depolarization
is due to \Ha-emitting gas at the velocity of the Perseus Arm.

\begin{figure}
\includegraphics[width=0.5\textwidth]{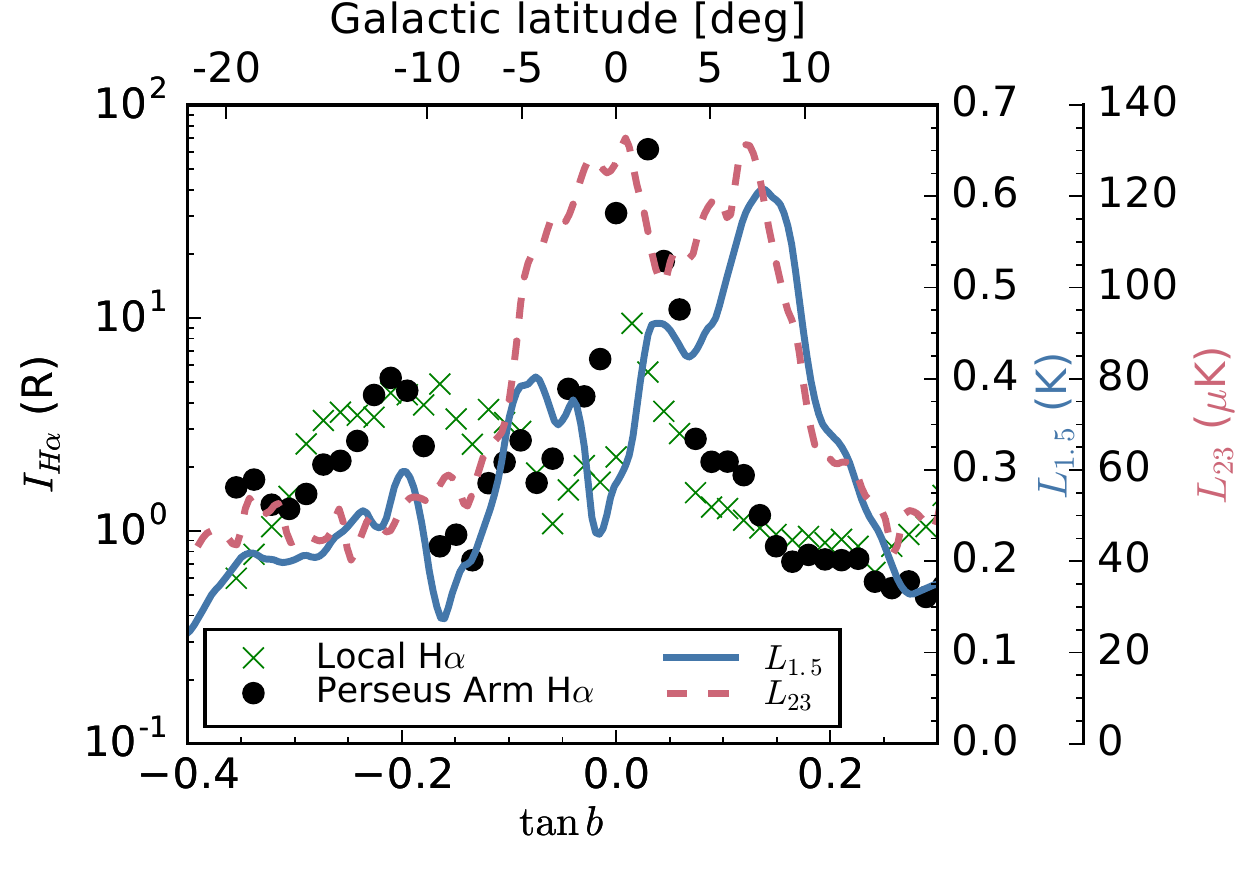}
\caption{Vertical profiles of polarized intensity and \Ha\ intensity (like Fig.~\ref{fig:depol_long}), averaged over $133\arcdeg < \ell < 136 \arcdeg$. Note that the \Ha\ is shown on a logarithmic scale, while the polarized intensity is shown on linear scales.}
\label{fig:pol_height}
\end{figure}

The vertical profile in Figure~\ref{fig:pol_height} shows that the $1.5 \GHz$
synchrotron intensity is highest at $b \approx +8\arcdeg$, with no sign of a
comparable enhancement at $b \approx -8\arcdeg$. The $23 \GHz$ polarized
emission is relatively bright over the range $-5\arcdeg \lesssim b \lesssim
+10\arcdeg$. The upper envelopes of $L_{1.5}$ and $L_{23}$ around $b = +10
\arcdeg$ are similar. The $1.5 \GHz$ intensity is much fainter at $b <
+5\arcdeg$, whereas the $23 \GHz$ intensity is comparable at $b \approx
0\arcdeg$ and $b=+8\arcdeg$. We interpret this as the result of beam depolarization;
$I_{\Ha} \approx 3-4 \R$ at $b=-5\arcdeg$, compared to $\approx 1
\R$ at corresponding latitudes above the plane. In our picture, the $23 \GHz$
and $1.5 \GHz$ emission each originate in the same volume. The $23 \GHz$
polarized intensity traces the full extent of the synchrotron-emitting region
without Faraday depolarization, while some of the $1.5 \GHz$ emission is
Faraday depolarized. 

\subsection{Using CGPS data to locate the Fan Region}
\label{sec:cgps}

\begin{figure*}
\includegraphics[width=0.9\textwidth]{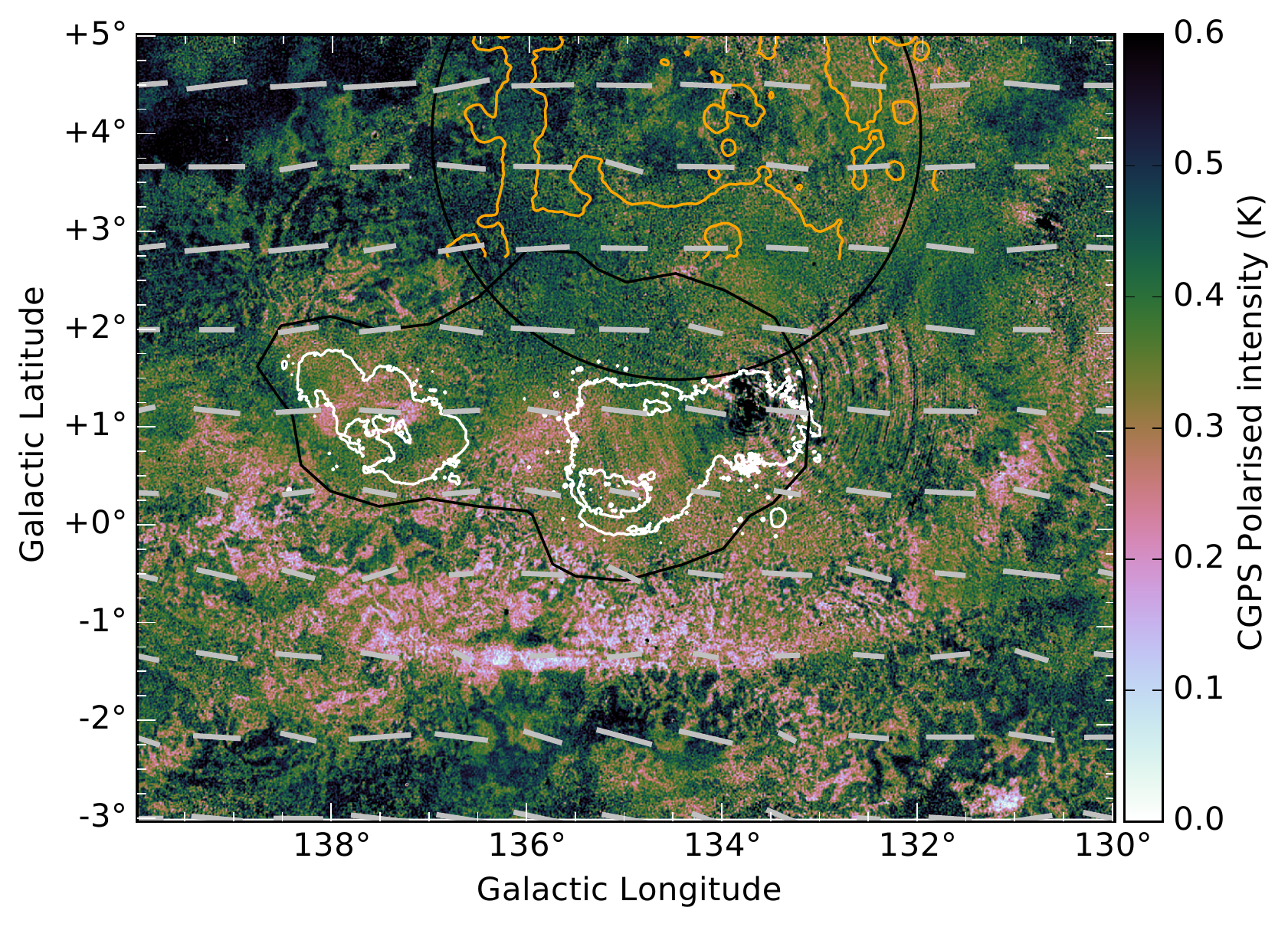}
\caption{Polarized intensity at $1.4 \GHz$ from the CGPS survey \citep{LandeckerReich:2010}. CGPS magnetic field vectors are overplotted, and white contours show the CGPS total intensity $I_{1.4}$ at $7.5 \K$. Orange contours show $4.8 \GHz$ total power in the W4 superbubble from \citet{GaoReich:2010,GaoReich:2015}. The black contour shows the WHAM-SS \Ha\ intensity at Perseus Arm velocities at $32 \R$ as in Fig.~\ref{fig:maps}$c$ and $f$. Note that the WHAM-SS data have a $1\arcdeg$ resolution and were obtained on a $1\arcdeg$ grid, so structure details on smaller scales are unreliable. The black circle is as in Figs.~\ref{fig:multifreq}--\ref{fig:maps}.}
\label{fig:cgps}
\end{figure*}

High resolution data are less subject to beam depolarization than the GMIMS
data. We use polarization data from the CGPS to obtain information relevant to
the Fan Region at low latitudes, shown in Figure~\ref{fig:cgps}. In the vicinity
of the Fan region the angular resolution is close to
$1 \arcmin$. We also refer to Figures $5-7$ of \citet{LandeckerReich:2010},
which show images of total intensity and polarized intensity from the CGPS over
$66\arcdeg < \ell < 186\arcdeg$.

The images show many areas of emisison that is structured on small angular
scales. Virtually all of this emission is unrelated to objects seen in total
intensity: these are typical Faraday screen structures.

The polarized intensity on and around W4 at $(134.7\arcdeg,+0.9\arcdeg)$ and W5
at $(137.6\arcdeg,+1.1\arcdeg)$, traced by $I_{1.4}$ (white) and \Ha\ (black)
contours in Figure~\ref{fig:cgps},
stands out from the highly structured surroundings. Toward these two \Hii\ regions
there is a marked absence of small-scale structure. Along the southern perimeter
of these \Hii\ regions, but offset from the edge, there is an abrupt transition to
highly structured polarized emission. This can also be seen clearly in Figures~7 and
8 of \citet{WestEnglish:2007}. \citet{GrayLandecker:1999} showed that this
transition below W4 is related to the \Hii\ region (and that is why the transition
mirrors its shape): there is an extended, low-density ionized halo that extends
beyond the boundaries defined by  $1.5 \GHz$ continuum imaging in total
intensity; this halo is evident with the WHAM data we show in
Figures~\ref{fig:maps} and \ref{fig:cgps} as well as in $2.7 \GHz$ total intensity
in Figure~7 of \citet{GrayLandecker:1999}. The gradient of Faraday rotation in
that envelope is evidently sufficient to cause beam
depolarization. The density of ionized gas drops off with distance beyond the \Hii\
region until the gradient of rotation is no longer sufficient to cause beam
depolarization, and the small-scale structure re-appears.

Our conclusion is that most, and probably all, of this highly
structured emission originates in the Perseus Arm, possibly in the vicinity of
W4 and W5 which are on the near surface of the Arm but possibly further inside
the Arm. The smoother emission seen towards W4 is more local emission. It is
expected that structures in the local emission will have an angular scale that
is larger than structures in the Perseus Arm.

There are two complications to this picture. First, the bright and compact
source W3 at $(133.8\arcdeg, +1.2\arcdeg)$ creates imaging artefacts, seen as rings
superimposed on W4. 
Second, there is a foreground object superimposed
on W5, evident in Figure~\ref{fig:cgps} as a pink ($L_{1.5} \approx 0.2 \K$) ring
at $(137.5\arcdeg, +1.1\arcdeg)$.
\citet{GrayLandecker:1998} demonstrate that this lens-like object is either an
enhancement of magnetic field or, more likely, an enhancement of electron
density, and conclude that it lies somewhere along the 2 kpc line of sight
between W5 and the Sun. 

Many \Hii\ regions other than W4 and W5 in the vicinity show similar behaviour: all of
them depolarize distant emission, leaving only foreground polarized emission
that has smooth structure \citep{LandeckerReich:2010}. This extends as far as
$\ell=173\arcdeg$, where the same depolarization effect is seen against the
G173+1.5 star formation complex in the Perseus Arm (distance $1.8 \kpc$;
\citealt{KangKoo:2012}).

The region LBN~0679 is a bright filament in total intensity that runs from
$(141\arcdeg,-1.5\arcdeg)$ to $(140\arcdeg,0\arcdeg)$. It is associated with \Hi\
at $-40 \pm 1.5 \kms$ \citep{Green:1989a}.
Other \Hii\ regions in the vicinity with similar velocities are at a distance
around $0.7 \kpc$ \citep{FosterBrunt:2015}. LBN~0679 shows the same polarization
signature: on the \Hii\ region the polarized intensity is smooth, while to either
side it is highy structured.

It is evident from Figures~\ref{fig:allsky} -- \ref{fig:maps}, \ref{fig:pol_height},
and \ref{fig:cgps} that there is strong depolarization in the plane at
1.5~GHz. There is an area of depolarization extending over
$128\arcdeg < \ell < 140\arcdeg, -1\arcdeg < b < +2\arcdeg$. In the CGPS data,
the southern edge of this depolarization zone has a very sharp edge, barely
resolved at arcminute resolution, evident in Figure~\ref{fig:cgps} as a curved
transition region from pink/white to green/black along $b = -1.5\arcdeg$. We refer
to this sharp edge as ``the Smile''; it is probably a shock front. Morphological
evidence suggests that the Smile is related to W4 and so is on the near side of
the Perseus Arm, although attempts to associate the Smile with any observed kinematic
features have failed \citep{LandeckerReich:2010}. In Figure~\ref{fig:pol_height},
we trace the polarized intensity at $\ell \approx 135\arcdeg$ as a function of $b$.
We see that $L_{1.5} \approx 0.4 \K$ at $b < -2\arcdeg$. At $b=-1.5\arcdeg$,
the polarized intensity drops abruptly to $L_{1.5} \approx 0.2 \K$. The polarized
intensity is similar from $-1.5\arcdeg \lesssim b \lesssim -0.5\arcdeg$ while
the structured appearance we discuss above is evident in Figure~\ref{fig:cgps}.
At $b \gtrsim -0.5\arcdeg$, the W4 \Ha\ and $2.7 \GHz$ emission becomes significant
and the polarized intensity increases to $L_{1.5} \approx 0.45 \K$.

We interpret the region in front of W4 and the Smile as follows. W4 completely
depolarizes all background emission by beam depolarization, even with the small
CGPS beam. The observed $\approx 0.45 \K$ emission in the direction of W4 is entirely
from the foreground and is therefore smooth. The Smile is outside the \Hii\
region and provides a much lower column of ionized gas than W4. The
depolarization in this region is therefore most likely Faraday screen depolarization
within the Smile. Because the emission
is more distant, it is more structured than the foreground emission observed in
front of W4. This interpretation is supported by the fact that there is no
depolarization evident in the Smile at $23 \GHz$; if the Smile were geometrical
depolarization, we would expect to see depolarization at $23 \GHz$.


Of particular interest is Sh2-202 around $(140\arcdeg,+2\arcdeg)$,
diameter $170 \arcmin$, at a distance of $0.97 \pm 0.08 \kpc$ \citep{FosterBrunt:2015}.
\citet{WilkinsonSmith:1974} noted that this \Hii\ region does not depolarize emission at
$610 \MHz$.\footnote{Note that there is a $10\arcdeg$ error in the longitude scale
of Figure~5 in \citet{WilkinsonSmith:1974}.} They took this as evidence that the Fan
Region emission arises at a distance less than that of Sh2-202. In the CGPS data
we see that Sh2-202 {\it{does}} leave an imprint on the $1.4 \GHz$ polarized
intensity image (Fig.~7 of \citealt{LandeckerReich:2010}),
so at least some of the polarized emission must originate beyond the distance of
Sh2-202. Again, the polarized emission across the face of Sh2-202 is smooth and
probably local in origin, while that from its surroundings has the more typical
mottled appearance that we identify with Perseus Arm emission.

Why can the influence of Sh2-202 as a Faraday screen be detected in the
polarization image, when Sh2-202 is not a bright emitter in total intensity?
The telescope is more sensitive to the Faraday rotation of a
volume of ionized gas than to its bremsstrahlung in the presence of a typical
interstellar magnetic field \citep{UyanikerLandecker:2003}. Based on the integrated
\Ha\ intensity of $37 \R$, we estimate $n_e = 1.6 \cucm$ in Sh2-202.
Assuming a line-of-sight magnetic field of $2 \uG$, this yields
$\phi \approx 100 \radmsq$. This produces a Faraday rotation in the \Hii\ region
at $1420 \MHz$ of $\Delta \psi \approx 250\arcdeg$, easily sufficient to cause
Faraday screen depolarization.

These results from the CGPS seem at first sight to be incompatible with the results
of \citet{WilkinsonSmith:1974}. However, their conclusions are based on a map of
the Fan Region at $610 \MHz$. In Section~\ref{sec:morphology}, we showed that the
appearance of the Fan Region below $1 \GHz$ is quite different from that at
$1.5 \GHz$ and above, with the low-frequency emission associated entirely with a nearby
feature.
 
\subsection{Comparison to the W4 superbubble} \label{sec:W4}

\begin{figure}
\includegraphics[width=0.5\textwidth]{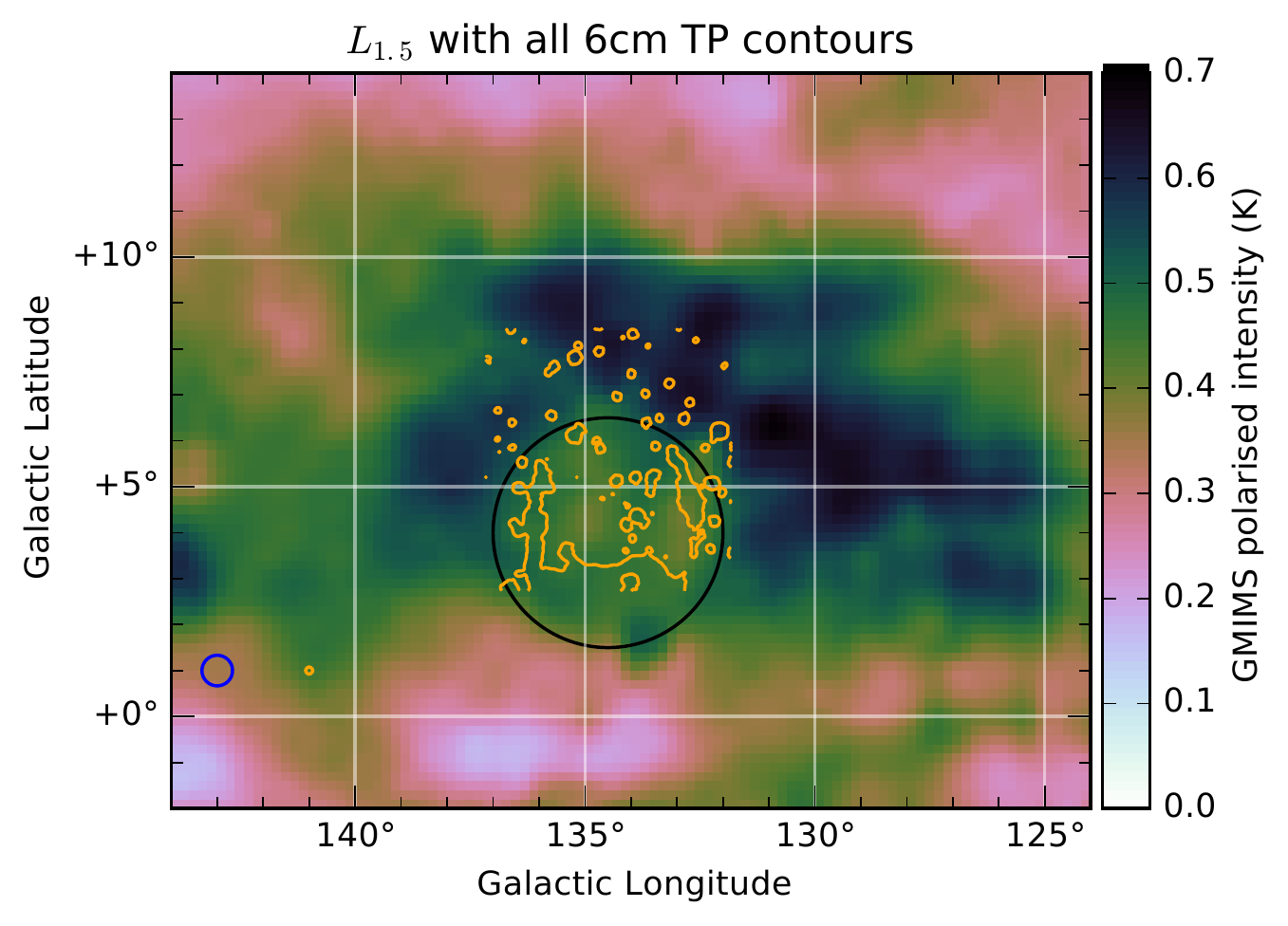}
\caption{1.5 GHz polarized continuum from GMIMS (as in Figs.~\ref{fig:multifreq}$b$ and \ref{fig:maps}) with orange contours of $4.8 \GHz$ total power in the W4 superbubble region from \citet{GaoReich:2010,GaoReich:2015} (as in Fig.~\ref{fig:cgps}). The beams of the $1.5 \GHz$ and $4.8 \GHz$ observations are shown in blue and orange, respectively, in the lower-left corner.}
\label{fig:maps6cm}
\end{figure}

Above W4 is the W4 ``chimney'' or superbubble \citep{NormandeauTaylor:1996}.
The superbubble walls are clearly seen in total intensity at $4.8 \GHz$ ($I_{4.8}$).
In Figure~\ref{fig:maps6cm} we superimpose $I_{4.8}$ contours on the map of $L_{1.5}$.
The $I_{4.8}$ contours fit inside the depression in $L_{1.5}$, strong
evidence that the W4 superbubble is responsible for the depolarization feature
centred at $(134.5\arcdeg, +4\arcdeg$). The relative locations of the wall seen
at $4.8 \GHz$ and the peak of the $1.5 \GHz$ polarized intensity are somewhat
uncertain due to the lower angular resolution of the GMIMS data. The outer edges
of the $I_{4.8}$ contours from the superbubble walls are shown as vertical orange lines in
Figure~\ref{fig:depol_long}. Again, it is clear that the $I_{4.8}$ superbubble walls
are inside the region of low $L_{1.5}$. The gradient in $L_{1.5}$ as a function
of longitude is much shallower than the gradient in $I_{4.8}$, likely at least partly
due to the much lower angular resolution of the $1.5 \GHz$ observations ($40'$ versus $9.5'$).
However, the separation between the peak of $L_{1.4}$ and the edge of the $I_{4.8}$
wall is $\approx 2\arcdeg$ on both sides of the superbubble, comfortably larger
than the resolution of the $1.5 \GHz$ observations. Therefore, the placement
of the superbubble wall inside, not coincident with, the reduced $L_{1.5}$
emission does not appear to be a resolution effect.
The total intensity maps at $1.4 \GHz$ \citep[hereafter GRR15]{WestEnglish:2007,GaoReich:2015}
show that the compressed walls of the superbubble (seen at $4.8 \GHz$) lie inside
a thicker envelope that is presumably ionized gas. The extent of this ionized
material probably determines the extent of the depolarization seen in the GMIMS data.
Since W4 is on the near side of the Perseus Arm, this is
clear evidence that at least some of the Fan Region emission must be generated in
the arm or beyond it.

\citetalias{GaoReich:2015} modeled the polarized emission along this line of sight
accounting for depth depolarization due to the W4 superbubble. They modeled
the superbubble as a shell structure with an inner radius of $65 \pc$
and an outer radius of $72 \pc$. The shell walls are evident as vertical structures
at $\ell \approx 136\arcdeg$ and $\ell \approx 133\arcdeg$ in Figure~\ref{fig:maps6cm}.
The GMIMS beam is $\approx 23 \pc$ at a distance of $2.0 \kpc$, so we do not resolve
the shell wall. The
\citet{WestEnglish:2007} DRAO Synthesis Telescope data (which lack zero-spacing
data and thus are not sensitive to degree-scale structure) show depolarization of
$0.025 \K$ in the shell walls relative to the $\approx 0.5 \K$ large-scale
polarized brightness temperature at $\lambda=21 \cm$. In Figure~\ref{fig:maps6cm},
the shell walls are {\em inside} the edge of the $0.7 \K$ Fan Region emission we
see with GMIMS. \citetalias{GaoReich:2015} concluded that the
depolarization by the shell walls is only $\approx 5\%$ at $21 \cm$.

Because the $\gtrsim 1\arcdeg$-scale structure in the $1.4 \GHz$ polarization images
presented by \citetalias{GaoReich:2015} is tied to the \citet{WollebenLandecker:2006}
survey (which does not show the large depolarization feature in the Fan Region, as
we discussed in Section~\ref{sec:morphology}), the lack of the large depolarization
feature in the data presented by \citetalias{GaoReich:2015} is expected. Our result is
therefore not inconsistent with \citetalias{GaoReich:2015}. The drop in intensity from
the Fan Region outside the shell wall ($L_{1.5} \approx 0.7 \K$) to inside the shell
($L_{1.5} \approx 0.5 \K$) is evident
with the GMIMS observations and indicates that the superbubble as a whole depolarizes
the Fan Region emission by $\approx 30\%$; this is the depolarization that is not
evident in the \citetalias{GaoReich:2015} data. Then the superbubble wall depolarizes
the $0.5 \K$ $1.4 \GHz$ emission on $\sim 10'$ scales by $\approx 5\%$; the
GMIMS data are not sensitive to variations in intensity on angular scales this small.


\section{Discussion}
\subsection{Polarization fraction} \label{sec:polfrac}

The Fan Region has exceptionally high polarized intensity, but the Stokes $I$ emission
is typical of the surrounding parts of the Galaxy, $I_{1.5} \approx 1.6 \K$. Equivalently,
the polarization fraction in the Fan Region is unusually high, $\approx 40\%$ (Fig.~\ref{fig:polfrac}).
Aside from the North Polar Spur, no other region of the sky has such strong emission
with such high fractional polarization. We have argued that
$\gtrsim 30\%$ of the $1.5 \GHz$ polarized emission originates in or beyond
the Perseus Arm through a morphological comparison of $L_{1.5}$ to \Ha\
and $I_{4.8}$ observations of ionized gas in the Perseus Arm
(Sections~\ref{sec:depol} -- \ref{sec:W4}).
However, we can argue for the same conclusion from the polarized radio continuum
data alone. \citet{BinghamShakeshaft:1967} were the first to do this, and their
argument, using modern data, is as follows.

The maximum possible polarized fraction of synchrotron
emission is $70\%$ \citep{GinzburgSyrovatskii:1965}, so in the simplest model,
$\sim 4/7$ of the radio continuum-emitting portion of the sightline contributes
to the Fan Region signal. (That estimate must, of course, take into account the varying
synchrotron emissivity along lines of sight through the Galaxy, as we do in
Section~\ref{sec:structure}.) Cosmic rays and magnetic fields, the ingredients
which produce synchrotron emission, are well-distributed around the Galaxy, so
the Stokes $I$ emission is produced along a long (at least several kpc) path.
If $4/7$ of the path contributes to the polarized emission, the polarized emission
must also be produced along a long path.

\citetalias{GaoReich:2015} applied the \citet{SunReich:2008} model of Galactic
synchrotron emission to the W4 (and Fan Region) sightline. This model incorporates
an enhanced synchrotron emissivity near the Sun \citep{FleishmanTokarev:1995}.
Figure~5 of \citetalias{GaoReich:2015} shows
that, in this model, $\approx 50\%$ of the Stokes $I$ emission originates within
$1 \kpc$ of the Sun and $\approx 25\%$ originates within $500 \pc$. This figure
describes the $4.8 \GHz$ emission, but assuming that the spectral index is constant
-- which we expect for Stokes $I$, which does not suffer from Faraday depolarization --
the same fractions should apply for $I_{1.5}$. If all of the Fan Region polarized
emission that reaches our telescope originates within
$500 \pc$ ($1 \kpc$) and the emitted fractional polarization is the maximal
$70\%$, the expected polarization fraction is therefore $18\%$ ($35\%$).
Therefore, the observed fractional polarization cannot be explained by emission
within $500 \pc$ of the Sun, and the fractional polarization in this model
within $1 \kpc$ of the Sun is still somewhat lower than (although probably within
the uncertainties of) the observed fractional polarization.

\subsection{Source of the synchrotron emission} \label{sec:emis}

\label{sec:warp}

What is the source of the emission? Synchrotron emission
requires cosmic rays and a component of the magnetic field perpendicular to the
line of sight. The scale heights of both cosmic rays and the regular component
of the magnetic field are highly uncertain but of order a few kpc
\citep{Ferriere:2001vr}, so there is no difficulty in finding the ingredients
required to generate synchrotron emission at moderate latitudes. In addition
to the morphological depolarization argument we presented in Section~\ref{sec:results},
there are then
two lines of argument which suggest that a significant component of the emission
is within the Perseus Arm or in the interarm region beyond the arm.

First, the Fan Region as seen at $1.5$, $23$, and $353 \GHz$ extends from
$b \approx -5\arcdeg$ to $b \approx +10\arcdeg$ (Section~\ref{sec:depol}
and Figs.~\ref{fig:allsky}, \ref{fig:multifreq}, and \ref{fig:pol_height}),
centred above the plane in a part of the Galaxy in which the emission beyond the
Perseus Arm is warped upward to $b \approx +5\arcdeg$ (Section~\ref{sec:features}
and Fig.~\ref{fig:b-v}). If the warp explains the asymmetry of the Fan Region
about the plane, some of the emission must be beyond the Perseus Arm.

Second, the magnetic field within the volume that generates the Fan emission must
be uniform to produce the highly ordered polarization signal.
In an interarm region the magnetic field is more likely to
remain coherent over the long path length required to produce uniform
synchrotron emission, and this may argue for at least a partial origin of the Fan
Region emission in regions beyond the Perseus Arm.
There is evidence for a more ordered field in interarm regions in the
face-on spirals M51 and NGC~6946 \citep{Beck:2007,FletcherBeck:2011}.

We therefore conclude that the most likely explanation for the Fan Region
emission is geometric. The Fan Region is the portion of the Galaxy in which
there is both a long path length with a coherent magnetic field, with a
significant component perpendicular to the line of sight, and a warp which allows
us to see much of the path length around depolarizing variations in the foreground
gas. In the next subsection, we construct a simple model applying this qualitative
discussion.

\subsection{Spiral structure and synchrotron emission} \label{sec:structure}

A number of authors have constructed models of the Galactic magnetic field aiming
to fit a number of observational constraints, including diffuse polarized emission.
Some mask the Fan Region from their fits \citep[e.g.][]{JanssonFarrar:2012},
while others ignore it, leaving high residuals \citep[e.g.][]{SunReich:2008,JaffeLeahy:2010,JaffeBanday:2011}. In general, these models
include a regular, planar field with a logarithmic spiral with a pitch angle of the
same sign as the spiral defining the gaseous and stellar spiral
structure\footnote{\citet{JanssonFarrar:2012} define this pitch angle sign as positive, while
\citet{SunReich:2008} define this sign as negative; we choose positive.} and a
value of $8\arcdeg \lesssim \pitch \lesssim 12\arcdeg$. \citet{Van-EckBrown:2011}
argue for an azimuthal field ($\pitch \approx 0\arcdeg$) in the outer Galaxy. The
models also typically include
turbulent, random, or striated fields; a vertical field; and contributions from
discrete structures. Although these models have had varying degrees of success
in matching the observed features in the Fan Region at low frequencies
($\nu \lesssim 2 \GHz$), none have fit it with low residuals as a
global feature. However, a spiral magnetic field with a positive pitch angle
places the maximal polarized intensity in the second quadrant ($\ell < 180\arcdeg$)
because the perpendicular component of the magnetic field is larger there than
at $\ell = 180\arcdeg$ or in the third quadrant.
Models developed in preparation for the Planck mission account for the $353 \GHz$
polarized emission from the Fan Region with only a global spiral magnetic field
and turbulence feature a maximum in polarization fraction around $\ell = 155\arcdeg$
\citep[Fig.~7 of][]{Miville-DeschenesYsard:2008}, in the second quadrant though
at a higher longitude than the Fan Region. These models too have high residuals
in polarization in the Fan Region
\citep{DelabrouilleBetoule:2013,Planck-CollaborationAde:2015,Planck-CollaborationAdam:2016a}.

We have attempted a very simple model which suggests one possible explanation while
illustrating the problems that still need to be solved before an adequate, complete
model of the Fan Region emission can be devised.
Our model (inspired by \citealt{JanssonFarrar:2012} and \citealt{SunReich:2008}) includes only
a radial variation in
the magnetic field, excluding reversals (which are irrelevant to the calculation
of synchrotron emissivity, which is proportional to $|\vec{B}|^2$), the vertical
component of the field, and separate spiral arms.
We assume that the magnetic field is
\newcommand{\rGC}{\ensuremath{r_\mathrm{GC}}}
\begin{equation} \label{eq:spiralB}
\vec{B} = \frac{B_0}{\rGC} \, \hat{b},
\end{equation}
where $\rGC$ is the distance from the Galactic Centre to the position, the unit vector along the logarithmic spiral defining the magnetic field is $\hat{b} = \sin{(\pitch)} \hat{r} + \cos{(\pitch)} \hat{\phi}$ in Galactocentric cylindrical coordinates, and the magnetic logarithmic spiral has a pitch angle $\pitch$. We further assume that the cosmic ray electron density is \citep[following][]{PageHinshaw:2007}
\begin{equation} \label{eq:ncre}
\ncre =  n_{\mathrm{cre},0} e^{-\rGC/h_r} \mathrm{sech}^2(z/h_z)
\end{equation}
where $h_r = 5 \kpc$ and $h_z = 1 \kpc$ are the scale length and height of the cosmic ray electron distribution. At $\rGC > 20 \kpc$, we set $\vec{B}$ and \ncre\ to zero. We model the thermal electron density outside the Solar circle as
\begin{eqnarray} \label{eq:n_e}
\nonumber
n_e &=& n_1 \frac{\mathrm{sech}^2(\rGC/A_1)}{\mathrm{sech}^2(r_\odot/A_1)} \exp \left(- \frac{|z|}{h_1} \right) 
+ \\ & & n_a \mathrm{sech}^2 \left(\frac{z}{h_a} \right) 
e^{-(s_\mathrm{Per}/w_a)^2} \mathrm{sech}^2 \left( \frac{\rGC-r_\odot}{2.0 \kpc} \right).
\end{eqnarray}
The first term is the smooth component of the \citet{TaylorCordes:1993} model of the Galactic electron density modified to use an exponential vertical distribution \citep{Schnitzeler:2012}. We choose $h_1 = 1.4 \kpc$ and $h_1 n_1 = 0.022 \kpc \cucm$, appropriate for the solar neighborhood \citep{GaenslerMadsen:2008,SavageWakker:2009},
and $A_1 = 20 \kpc$. The second term represents the contribution from the Perseus spiral arm. We use the \citet{ReidMenten:2014} definition of the arm as a logarithmic spiral (Fig.~\ref{fig:arms}); $s_\mathrm{Per}$ is the distance from a given position to the nearest point in the spiral arm, $n_a = 0.084 \cucm$ is the midplane electron density in the arm, and $w_a = 300 \pc$ is the assumed scale length of the electron density in the arm \citep{TaylorCordes:1993}.

For each longitude, we calculate the polarized synchrotron intensity from the polarization vector \citep{Burn:1966ug,SokoloffBykov:1998,OSullivanBrown:2012}
\begin{equation} \label{eq:P}
\mathcal{P} = \int^{\mathrm{observer}}_{\mathrm{back}} \varepsilon \, \ncre(s) \, \left(\vec{B}(s) \times \vec{\hat{s}}\right)^2 \, e^{2i\psi(s,\lambda)} ds 
\end{equation}
where the polarization angle is $\psi(s,\lambda) = \psi_0(s) + \phi(s) \lambda^2$ and
the Faraday depth is defined in equation~(\ref{eq:phi}). We use 
equations~\ref{eq:spiralB}--\ref{eq:n_e} to determine $n_\mathrm{cre}$, $n_e$, and
$\vec{B}$. We choose the emissivity $\varepsilon$ such that
$n_\mathrm{cre,0} \varepsilon = 0.7 \K \cucm$ because this produces a synchrotron
intensity in the Fan Region of $\approx 0.5 \K$, as observed, but the units of the
output intensity can be scaled arbitrarily.
We stay in the Galactic plane ($z=0$). This implicitly but inexactly
accounts for the warp: because the brightest $1.5 \GHz$ emission in the Fan Region has a
non-zero latitude ($b \approx +8\arcdeg$), the altitude probed by the sightline
would increase with distance as $d \sin b$. In a warped
disk, the latitude of the midplane increases with distance, so the distance between
the sightline and the midplane is less than $d \sin b$.
The inclusion of equations~\ref{eq:phi} and \ref{eq:n_e} accounts for depth depolarization
and equation~\ref{eq:P} includes geometrical depolarization, but
we leave the inclusion of beam depolarization for Paper~II because it involves
a number of additional assumptions that are more difficult to quantify.

\begin{figure}
\includegraphics[width=0.5\textwidth]{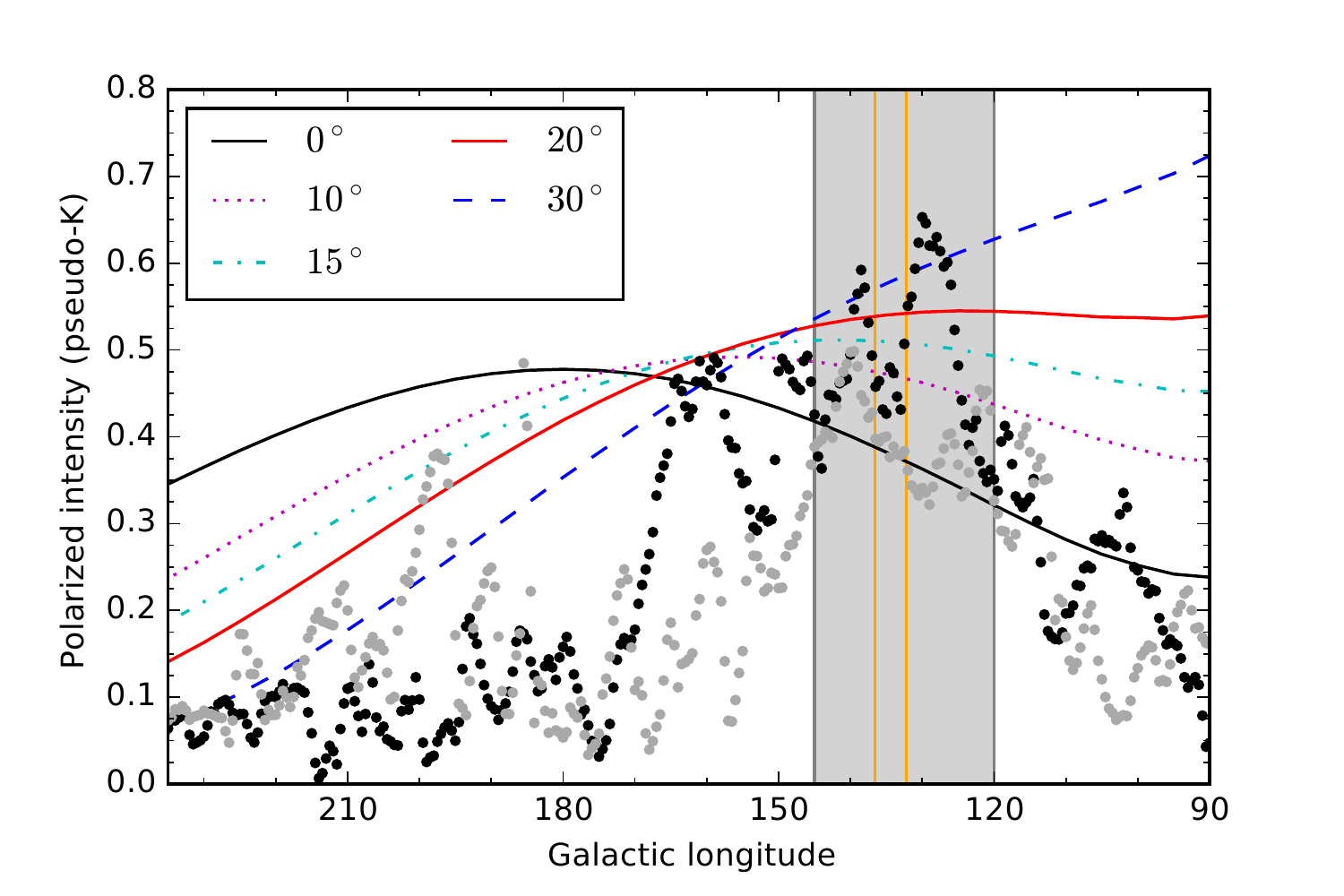}
\caption{Synthetic polarized synchrotron intensity at $\lambda=20\cm$ as a function of longitude for a
range of magnetic spiral pitch angles $\pitch$ as described in Section~\ref{sec:structure}.
This model includes depth and geometrical but not beam depolarization. As in Fig.~\ref{fig:arms}, the
longitude range of the Fan Region is shaded grey. We also show polarized intensity
from GMIMS-HBN as a function of Galactic longitude
at $b = +4.5\arcdeg$ (black dots) and $b=-4.5\arcdeg$ (light gray dots).
The orange lines denote the outer edge of the W4 superbubble as in
Fig.~\ref{fig:depol_long}.} 
\label{fig:synch_l}
\end{figure}

We show this modeled midplane polarized synchrotron intensity as a function of Galactic
longitude in the outer Galaxy in Figure~\ref{fig:synch_l}. The intensity is
maximal at $\ell=180\arcdeg$ for $\pitch=0$ (a circular field). For non-zero
pitch angles, the polarized synchrotron intensity is much higher in the second quadrant
($\ell < 180\arcdeg$) than in the third quadrant ($\ell > 180\arcdeg$) due to
the small perpendicular component of $\vec{B}$ in the third quadrant. In the
second quadrant, absent depth depolarization effects, the polarized synchrotron intensity
is relatively flat as a function of longitude when $\pitch > 0$. However, due to
depth depolarization, the observed polarized synchrotron intensity reaches a maximum and
then decreases towards lower longitude. The longitude of the peak is lower for
higher \pitch; for $15\arcdeg \lesssim \pitch \lesssim 20\arcdeg$, the intensity
peaks in the Fan Region. For $\pitch=30\arcdeg$, the intensity does not peak in the
second quadrant; it increases monotonically from $\ell=180\arcdeg$
to $\ell=90\arcdeg$.

We also show polarized intensity as a function of longitude in the entire portion
of the outer Galaxy observed with GMIMS-HBN in Figure~\ref{fig:synch_l}.
We show $L_{1.5}$ at $b = +4.5\arcdeg$, which runs through the centre of the Fan
Region, and $b=-4.5\arcdeg$, which runs through the brightest polarized emission
in the third quadrant (see Fig.~\ref{fig:allsky}$a$). The longitude of peak
emission from the $\pitch \approx 20\arcdeg$ models roughly matches that in the
observations. However, the observed polarized intensity falls off much more
rapidly on either side of the peak than these models but more slowly than the
$\pitch=0\arcdeg$ model. In particular, the modeled $\pitch = 20\arcdeg$ polarized
intensity is close to flat at $\ell < 130\arcdeg$.

The addition of a source of cosmic ray electrons concentrated around W4 can produce a peak
in polarized intensity that roughly matches the observed cross-section through the
Fan Region shown in Figure~\ref{fig:synch_l}, but (a) there is no observational
evidence, in the way of supernova remnants, for any source of high-energy electrons,
and (b) this extra synchrotron emissivity would produce a peak in total intensity
as well, which is not observed. The puzzling feature of the Fan Region is its
exceptionally high polarized intensity, not its total intensity. The remaining
difficulty is to explain the very high level of uniformity of the magnetic field
implied by the high fractional polarization, and to explain that field regularity
over the long path length that is implied by the observational evidence presented
in this paper.

None of these simple models accurately match the morphology of the Fan Region.
Models with a pitch angle $\pitch \approx 10\arcdeg$, as others have generally
preferred \citep[e.g.][]{SunReich:2008,Miville-DeschenesYsard:2008,JanssonFarrar:2012}
produce a peak in intensity at higher longitudes than is observed in the Fan Region,
while the steep pitch angle which best matches the longitude of peak intensity produces
a higher intensity than is observed to lower longitudes than the Fan Region.


We conclude that it
is unlikely that any simple model of the Galactic magnetic field will explain all
of the observations. However, a spiral magnetic field with a relatively-steep
pitch angle in the outer Galaxy can plausibly explain a distant origin of
$\approx 30\%$ of the Fan Region emission. The Fan Region is in the quadrant of
the Galaxy one would expect for a feature that arises due to synchrotron emission
from the Galactic magnetic field with a spiral with a positive pitch angle. We note
that the gas in the Outer Arm has a steep pitch angle, $\pitch = 18.6\arcdeg$
\citep[Fig.~\ref{fig:arms} and][]{ReidMenten:2014}.
It is possible that the Fan
Region originates in a part of the Galaxy with a steeper pitch than is preferred
at $r_{\mathrm{GC}} \lesssim 10 \kpc$, although this result is inconsistent with
the azimuthal field toward the anticentre preferred by \citet{Van-EckBrown:2011}
based on RM measurements of extragalactic sources.

\section{Summary and conclusions} \label{sec:conclusions}

We have used new GMIMS-HBN observations and other published observations to describe
the morphology of polarized continuum emission found in data over $0.4 - 353 \GHz$
in the Fan Region. In summary, our key observational findings are:
\begin{enumerate}
\item All-sky maps of $1.5$, $23$, and $353 \GHz$ polarized emission from the Fan Region show that the Fan Region is roughly coincident with \Ha\ emission from the Perseus Arm, especially the \Ha\ emission from around the W3/W4/W5 complex of \Hii\ regions. They are similar in both location and angular extent on $\approx 10\arcdeg$ scales (Figs.~\ref{fig:allsky} and \ref{fig:multifreq}$a$, $b$).
\item While the large-scale {\em extent} of the Fan Region polarized emission at
$> 1 \GHz$ roughly coincides with the Perseus Arm \Ha\ emission, the detailed
structure on $\approx 1\arcdeg$ scales shows anticorrelation.
A morphological comparison of GMIMS-HBN data at $1.5 \GHz$ to \Ha\ (Figs.~\ref{fig:maps} -- \ref{fig:depol_long} and Section~\ref{sec:depol}) and radio continuum total intensity observations with high angular resolution (Figs.~\ref{fig:cgps} and \ref{fig:maps6cm} and Section~\ref{sec:cgps}) shows that depolarization evident in polarized intensity is correlated with bright features in ionized gas related to the W4 star formation region and superbubble.
The polarized intensity $L_{1.5}$ is $30\%$ lower in regions with high integrated \Ha\ intensity from the Perseus Arm ($I_{\Ha} > 5 \R$) than in regions with low integrated \Ha\ intensity from the Perseus Arm ($I_{\Ha} < 3 \R$; Fig.~\ref{fig:depol_ha}).
\item At frequencies lower than $1.5 \GHz$, the size of the Fan Region decreases
with decreasing frequency. At $\nu \lesssim 600 \MHz$, the morphology of the Fan Region is quite different than at $\nu \gtrsim 1 \GHz$: the low-frequency emission is a ring centred at $(137\arcdeg, +8\arcdeg$), while the high-frequency emission extends to significantly lower longitudes ($\ell \approx 115\arcdeg$) and latitudes ($b \approx -5\arcdeg$) and does not have a ring-like component (Section~\ref{sec:morphology} and Fig.~\ref{fig:multifreq}).
\item The fractional polarization of the parts of the Fan Region with bright
polarized emission at $1.5 \GHz$ is high, $L_{1.5}/I_{1.5} \approx 40\%$
(Section~\ref{sec:morphology} and Fig.~\ref{fig:polfrac}).
\end{enumerate}

Observational fact (ii) leads us to conclude that at least $30\%$ of the $1.5 \GHz$ polarized continuum emission seen in the brightest parts of the Fan Region originates in or beyond the Perseus Arm. If the Perseus Arm acts as a \citet{Burn:1966ug} slab which depolarizes all background emission -- a conclusion supported by the smoothness of the CGPS data -- most or all of the remaining $\approx 70\%$ would originate in front of the Perseus Arm. Observational fact (iii) suggests that this high-frequency structure is different in physical origin than most of the low-frequency ($\nu \lesssim 600 \MHz$) emission that was first associated in the literature with the Fan Region.
The more distant high-frequency emission is likely more depolarized at low frequencies (which we will discuss more in Paper~II), so the emission and Faraday effects in local features, within about $500 \pc$, may dominate. These local features appear to make an insignificant contribution to the Fan Region at high frequencies.
The rest of our conclusions apply to the $\nu > 1 \GHz$ data.

Observational fact (iv) implies that the entire line of sight must be involved in
generating the Fan Region synchrotron emission (Section~\ref{sec:emis}).
Even though a majority (up to $70\%$) of the $1.5 \GHz$ emission could originate in front of the Perseus Arm, it seems highly unlikely that most of the foreground emission
is within $\approx 500 \pc$ while $30\%$ of the emission originates in or
beyond the Perseus Arm.
It is far more plausible that the origin of the Fan Region is not a discrete, local structure. Therefore, the Fan Region must be a very large phenomenon, several kpc in extent, which we can only explain as a consequence of Galactic structure and geometry. We cannot confine it, as most previous authors have done, to the nearest $500 \pc$. 


We could have reached many of our conclusions about the distance of the Fan Region
without the GMIMS data or the \Ha\ data based entirely on the high fractional polarization,
the modelling of Section~\ref{sec:structure}, and the evidence from the {\em Planck} $353 \GHz$
data, together with the modelling done by the Planck consortium (see references
in Section~\ref{sec:structure}). The GMIMS and WHAM data, taken together,
reinforce our conclusion.

Our determination of the distance to the Fan Region emission from the GMIMS and
WHAM data is strongly supported by the correlation of the
$\approx 4\arcdeg$-diameter region of reduced polarized intensity around
$(134.5\arcdeg, +4\arcdeg)$ with \Ha\ intensity at Perseus Arm velocities
and the W4 superbubble.
This is a relatively small patch of sky.
Our extrapolation from the Fan Region to an analysis of the Galactic magnetic field rests
on analysis of observations over much larger scales, the fact that the polarized
intensity is much higher in the second quadrant (where the Fan Region extends
over $\sim 60\arcdeg$) than in the third quadrant.


We suggest three ideas which explain some of the observed features of the Fan Region
emission:
\begin{enumerate}
\item A spiral magnetic field with a steep pitch angle ($\pitch \approx 15\arcdeg-20\arcdeg$)
moves the longitude of peak emission to $\ell \approx 130\arcdeg$, in the Fan Region
(Fig.~\ref{fig:synch_l}). However, the increase in intensity is less sharp than
observed, and the pitch angle is significantly steeper than the $\pitch \approx 10\arcdeg$
preferred by most existing models.
\item An increase in the synchrotron emissivity associated with W4 could be scaled
to produce the observed Fan Region intensity. Qualitatively, this is consistent
with the morphology of the brightest emission from the Fan Region, surrounding
W4. However, the $60\arcdeg \times 30\arcdeg$ extent of the Fan Region is much
larger than W4 itself. Moreover, it is not obvious how to
increase the polarized intensity without also increasing the total intensity,
and it is also not obvious how increased intensity in a presumably-turbulent region
associated with star formation could produce such regular polarization vectors and
such a high polarization fraction.
This idea also does not explain the offset of the Fan Region above $b=0\arcdeg$.
\item Due to the warp, distant ($d > 2 \kpc$) portions of this part of the Galaxy
are centred at $b = +3\arcdeg$ to $+9\arcdeg$ (Fig.~\ref{fig:b-v}). Moreover, the
pitch of the gaseous arms is considerably steeper in the outer Galaxy
(Fig.~\ref{fig:arms}); because most models of the magnetic field are not constrained
so far out ($R \gtrsim 20 \kpc$), a steeper pitch angle may be consistent with
existing models. It is not clear that the cosmic ray electron density is high
enough in the outer Galaxy, where there is little star formation, to produce the
observed synchrotron intensity. However, in M51, there is detectable polarized
emission at $1.4 \GHz$ out to $\sim 50 \kpc$ from the galactic centre, well
beyond the optical spiral arms \citep{FletcherBeck:2011}.
\end{enumerate}

None of these three ideas explain all of the observed features of the Fan Region,
so the origin of the Fan Region remains puzzling. We conclude that $\approx 30\%$
of the integrated $1.5 \GHz$ Fan Region emission is depolarized by ionized gas
in the Perseus Arm, suggesting that it is a puzzling Galactic-scale feature, not a
relatively-small, purely local feature. This result suggests that future detailed models
of the Galactic
magnetic field should attempt to fit the Fan Region with the prior that the
emission originates along a long path length or at a large distance, perhaps
incorporating the warp or a spiral magnetic field with a large pitch angle in at
least part of the Galaxy.

%

\section*{Acknowledgements}

ASH acknowledges useful discussions with R.\ A.\ Benjamin and V.\ Jelic.
TLL acknowledges useful discussions with T.\ Foster and J.\ C.\ Brown. We thank
two referees for comments which strengthened the paper.

This research made use of APLpy\footnote{\url{http://aplpy.github.com}} and
Astropy \citep{Astropy-CollaborationRobitaille:2013}.

The Wisconsin H-Alpha Mapper is funded
by the US National Science Foundation. The Dominion Radio Astrophysical
Observatory is operated as a National Facility by the National Research Council Canada.
ASH was partially supported by NSF grant AST-1442650.
KD and MW were supported by the Natural Sciences and Engineering Research Council of Canada.
BMG acknowledges the support of the Australian Research Council through grant FL100100114. The Dunlap Institute is funded through an endowment established by the David Dunlap family and the University of Toronto.
NMM-G acknowledges the support of the Australian Research Council through Future Fellowship FT150100024.
MH acknowledges the support of research program 639.042.915, which is partly
financed by the Netherlands Organization for Scientific Research (NWO).




\bibliographystyle{mnras}
\bibliography{bibdesk_bibtex} 




%
%


\bsp	
\label{lastpage}
\end{document}